\pdfoutput=1

\documentclass[11pt]{article}

\usepackage[]{ACL2023}

\usepackage{times}
\usepackage{latexsym}
\usepackage{amsmath}
\usepackage{algorithm, algorithmic}
\usepackage{multicol}
\usepackage{booktabs}
\usepackage{multirow}
\usepackage[T1]{fontenc}
\usepackage{xspace}
\usepackage{graphicx}

\usepackage{tablefootnote}
\newcommand*{\AlgName}{\text{H-STAR}\@\xspace}

\usepackage[T1]{fontenc}

\usepackage[utf8]{inputenc}

\usepackage{microtype}

\usepackage{inconsolata}

\title{\textcolor{black}{H-STAR:} LLM-driven \textcolor{black}{H}ybrid \textcolor{black}{S}QL-\textcolor{black}{T}ext \textcolor{black}{A}daptive \textcolor{black}{R}easoning on Tables}

\author{Nikhil Abhyankar\textsuperscript{\tiny 1}, Vivek Gupta\textsuperscript{\tiny 2}, Dan Roth\textsuperscript{\tiny 3}, Chandan K. Reddy\textsuperscript{\tiny 1} \\
        \textsuperscript{\tiny 1}Virginia Tech, \textsuperscript{\tiny 2}Arizona State University, \textsuperscript{\tiny 3}University of Pennsylvania \\
        \tt \small {nikhilsa@vt.edu, vgupt140@asu.edu, danroth@seas.upenn.edu, reddy@cs.vt.edu}}

\begin{document}
\maketitle
\begin{abstract}
Tabular reasoning involves interpreting natural language queries about tabular data, which presents a unique challenge of combining language understanding with structured data analysis. Existing methods employ either textual reasoning, which excels in semantic interpretation but struggles with mathematical operations, or symbolic reasoning, which handles computations well but lacks semantic understanding. This paper introduces a novel algorithm \AlgName that integrates both symbolic and semantic (textual) approaches in a two-stage process to address these limitations. \AlgName employs: (1) step-wise table extraction using `multi-view' column retrieval followed by row extraction, and (2) adaptive reasoning that adapts reasoning strategies based on question types, utilizing semantic reasoning for direct lookup and complex lexical queries while augmenting textual reasoning with symbolic reasoning support for quantitative and logical tasks. Our extensive experiments demonstrate that \AlgName significantly outperforms state-of-the-art methods across three tabular question-answering (QA) and fact-verification datasets, underscoring its effectiveness and efficiency.\footnote{The code and data are available at: \url{https://github.com/nikhilsab/H-STAR}}
\end{abstract}

\section{Introduction}
\begin{figure}[!htbp]
    \small
    \centering
    \includegraphics[width=1\linewidth]{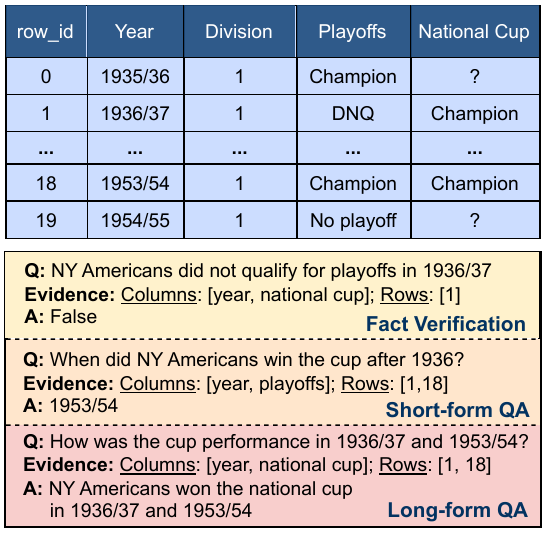}
    \caption{\small An illustration of different tabular reasoning tasks \textcolor{black}{ \bf (a) Fact-verification}, \textcolor{black}{\bf (b) Short-form QA}, and \textcolor{black}{\bf (c) Long-form QA}. For each task, question Q is paired with its answer A, which varies by task. Evidence shows the relevant columns and rows needed to answer the question.
    }
    \vspace{-0.5em}
    \label{fig:enter-label}
\end{figure}

Tabular data are among the most widely used formats for storing structured information in real-world applications. Tabular reasoning presents an inherently challenging problem, requiring logical, mathematical, and textual reasoning over unstructured queries and structured tables \cite{ye2023large}. Thus, understanding and inferring tabular data has become a significant area of research in natural language processing. Tabular reasoning tasks (Figure \ref{fig:enter-label}), such as table-based question answering \cite{pasupat2015compositional, nan2022fetaqa} and table-based fact verification \cite{chen2019tabfact}, have been extensively explored in the past.

\begin{figure*}[!htbp]
\centering
\includegraphics[width=1\linewidth]{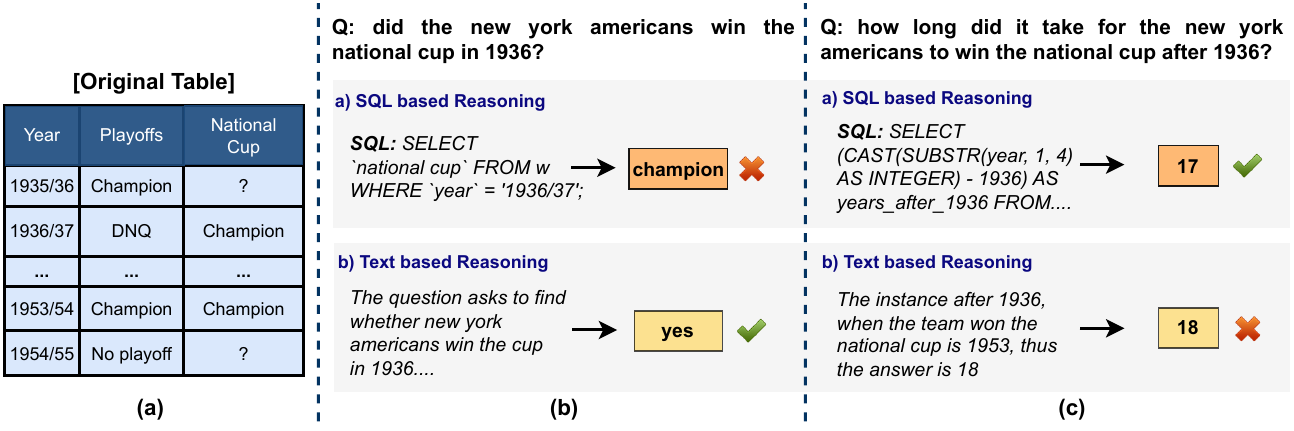}
  \vspace{-0.5em}
  \caption {\small \textbf{An illustration highlighting the complexity of table reasoning and the need for an integrated approach}: (a) Original table, (b) Symbolic reasoning misinterprets the question and returns a value instead of a yes/no response, and (c) Text-based approach fails to solve a math question correctly, leading to an incorrect answer.}
  \vspace{-.5em}
  \label{fig: hard_example}
\end{figure*}
\textcolor{black}{Advancements in Large Language Models (LLMs) have led to better performance across various tasks through carefully crafted prompts. In the domain of table reasoning, symbolic approaches such as {Program of Thought} \cite{chen2023program} and textual methods such as Chain of Thought (CoT) \cite{wei2022chain, brown2020language} have been explored. However, when applied individually (see Figure \ref{fig: hard_example}) these methods often struggle due to the complexities imposed by the intricate mix of numerical, temporal, and textual data, coupled with complex table structures \cite{shi2023large, liu2023rethinking, sui2024table}. Textual reasoning excels in natural language understanding but often misinterprets table structures and struggles with quantitative reasoning. In contrast, SQL-based approaches are strong in quantitative problem solving, but perform poorly on noisy or unstructured inputs \cite{liu2023rethinking}. As a result, recent techniques \cite{cheng2022binding, ye2023large, wang2023chain, nahid2024tabsqlify} that rely solely on textual or symbolic reasoning suffer from similar limitations. Therefore, it is crucial to explore the question: \emph{How can \textbf{symbolic and textual approaches be integrated} into a hybrid method for tabular reasoning that \textbf{leverages their complementary strengths while mitigating their individual limitations}?}}

In this paper, we introduce \AlgName which combines symbolic and textual reasoning abilities of LLMs, achieving the \emph{\textit{best of both worlds}}. \AlgName decomposes the table reasoning task into two sub-tasks: \textbf{(a) Table Extraction} and \textbf{(b) Adaptive Reasoning}. In the table extraction phase, we employ \textcolor{black}{a step-by-step `multi-view' chain, first identifying relevant columns using the original table and its transposed form. The filtered table is then used for row extraction, contrary to other approaches, which use the original table for both row and column selection \cite{ye2023large, nahid2024tabsqlify}.} Focusing on relevant table cells helps LLMs by providing the right context, thus reducing hallucinations in reasoning. We employ a combination of SQL-based and text-based techniques in \AlgName's table extraction phase. In the adaptive reasoning phase, we use the language comprehension capabilities of LLMs to guide them with few-shot examples to decide when to support semantic methods with symbolic (SQL) approaches. \AlgName employs semantic reasoning universally, using it exclusively for direct lookup, common sense, and complex lexical queries while using an additional SQL step for quantitative, mathematical, and logical tasks. This strategy optimizes performance on diverse question types, supporting text comprehension with computation when required.

\textcolor{black}{We demonstrate the efficacy of \AlgName across three tabular benchmarks involving table QA and table fact verification. Our experiments show that using the combined approach of \AlgName, aligned with multi-view extraction and adaptive reasoning outperforms all prior state-of-the-art approaches. Our analysis demonstrates that \AlgName is robust and generalizable across LLMs, showcasing its superior reasoning capabilities across various tasks. Our main contributions are as follows:}

\vspace{0.25em}
\noindent (a) We introduce \AlgName, a two-step method unifying table extraction with adaptive reasoning to enhance performance on tabular reasoning. It is a hybrid approach that effectively integrates symbolic (SQL logic) and semantic (textual) methods.

\vspace{0.25em}
\noindent (b) Through extensive experiments, we show that \AlgName's hybrid approach surpasses state-of-the-art methods that rely solely on either symbolic or semantic techniques, especially when reasoning over larger tables.

\vspace{0.25em}
\noindent (c) Our ablation study and analysis highlight \AlgName's effectiveness in table extraction and reasoning across various models and datasets, underscoring the significance of each stage in enhancing overall performance. 
\vspace{0.25em}

\noindent The code and additional resources are available at: \url{https://hstar-llm.github.io/}.
\section{\AlgName Approach}
\AlgName decomposes the table reasoning task into extraction and reasoning stages, combining LLM's textual comprehension with symbolic reasoning. It converts the original table ({\fontfamily{qcr}\selectfont T}) into a query-specific table ({\fontfamily{qcr}\selectfont T\textsubscript{CR}}) (refer Algorithm \ref{alg:cap}). Figure \ref{fig:main_pipeline} illustrates our \AlgName approach. Unlike DATER \cite{ye2023large} and TabQSLify \cite{nahid2024tabsqlify}, which use textual reasoning and text-to-SQL respectively, \AlgName integrates both reasoning types in a complementary manner for table extraction. \textcolor{black}{Our reasoning step dynamically uses symbolic techniques alongside semantic methods} for quantitative questions, overcoming the limits of pure textual reasoning. \AlgName operates in two main stages: {\it 1) Table Extraction} and {\it 2) Adaptive Reasoning}. Implementation details, including prompts, input table formats, and hyperparameters, are provided in Appendix \ref{sec:input_format} and \ref{sec: implm_details}.

\subsection{Table Extraction}
\begin{figure*}[!htbp]
  \centering
  \includegraphics[width=0.99\linewidth]{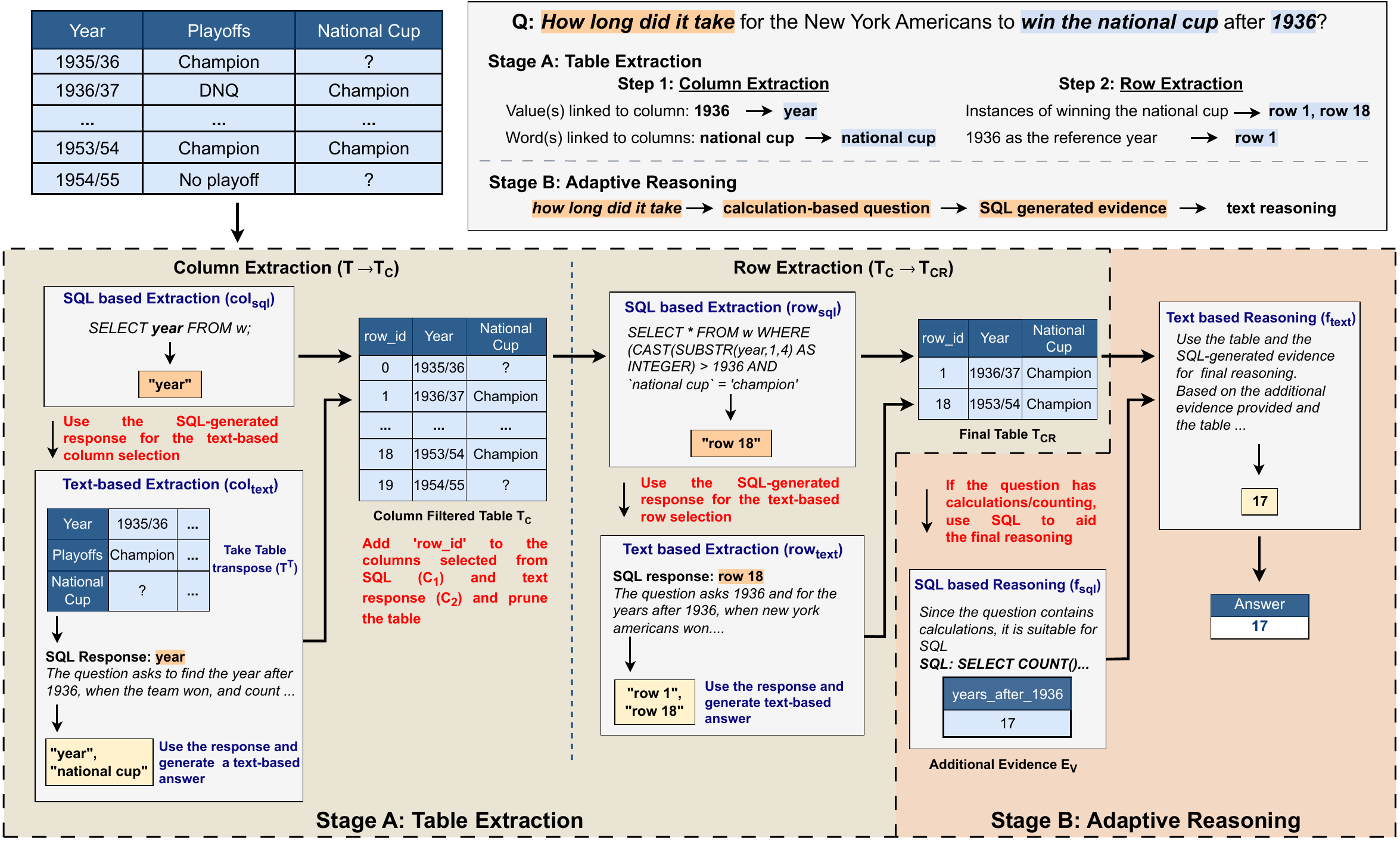}
  \caption {\small \textbf{An overview of \AlgName}, consisting of a combination of code generation and text-based verification. Given a complex table and its question, \AlgName answers using (a) \textbf{Table Extraction:} extracts the question-specific table from the original by first selecting the columns followed by rows. (b) \textbf{Adaptive Reasoning}: when the question has any mathematical component, it generates an additional table using SQL used in the textual reasoning step.}
  \vspace{-1.em}
  \label{fig:main_pipeline}
\end{figure*}

Two-dimensional tables consist of columns and rows. Table extraction involves a two-step reasoning chain: \textbf{(1)} column extraction and \textbf{(2)} row extraction (see Figure \ref{fig:main_pipeline}). As illustrated in Algorithm \ref{alg:cap}, for a given table {\fontfamily{qcr}\selectfont T}, table extraction returns a table {\fontfamily{qcr}\selectfont T\textsubscript{CR}} (lines 1-8) based on the input query {\fontfamily{qcr}\selectfont Q}.

\paragraph{Column Extraction.}
As shown in Algorithm \ref{alg:cap}, \AlgName uses a `multi-view' technique, first using the original table {\fontfamily{qcr}\selectfont T} (line 1) followed by its transposed form {\fontfamily{qcr}\selectfont T\textsuperscript{T}} (line 2) for column extraction. \textcolor{black}{Changing the `view' acts as a verifier accounting for any information missed due to the table structure, thus improving the overall performance (see Appendix \ref{sec:transposed}).} In Figure \ref{fig:main_pipeline}, given an input table {\fontfamily{qcr}\selectfont T} and the question {\fontfamily{qcr}\selectfont Q} ({\it `How long did it take the New York Americans to win the National Cup after 1936?'}), the LLM employs {\fontfamily{qcr}\selectfont col\textsubscript{sql}} to generate an SQL query to extract columns {\it `year'} ({\fontfamily{qcr}\selectfont C\textsubscript{1}}). This is followed by the text-based step {\fontfamily{qcr}\selectfont col\textsubscript{text}} which processes table {\fontfamily{qcr}\selectfont T\textsuperscript{T}} to return columns {\fontfamily{qcr}\selectfont C\textsubscript{2}} ({\it `year', `national cup'}). This step captures the `national cup' information relevant to answering the query, which is missed by {\fontfamily{qcr}\selectfont col\textsubscript{sql}}. Table {\fontfamily{qcr}\selectfont T} is then filtered for columns {\it `year', `national cup'} ({\fontfamily{qcr}\selectfont C'}) derived from {\fontfamily{qcr}\selectfont C\textsubscript{1}} and {\fontfamily{qcr}\selectfont C\textsubscript{2}} to obtain table {\fontfamily{qcr}\selectfont T\textsubscript{c}} (lines 3,4).

\vspace{0.5em}
\begin{algorithm}[!htbp]
\setlength{\intextsep}{0pt}
\small
\caption{\textbf{\AlgName}}
\label{alg:cap}
\begin{algorithmic}[1]
\REQUIRE (T, Q) $\rightarrow$ Table - Question pair
\ENSURE $\hat{A}$ $\rightarrow$ Predicted Answer\\
\textbf{Stage A: Table Extraction}\hfill \\
\underline{Column Extraction} \hfill
\STATE \quad C\textsubscript{1}$\leftarrow$ {\fontfamily{qcr}\selectfont col\textsubscript{sql}}(T, Q) \hfill {\small{$\triangleright$ \textcolor{black}{SQL approach}}
\STATE \quad C\textsubscript{2} $\leftarrow$ {\fontfamily{qcr}\selectfont col\textsubscript{text}}(\textsc{T}\textsuperscript{T}, \textsc{Q})
\hfill {\small{$\triangleright$ \textcolor{black}{Text approach}} \\$\triangleright$ $\textsc{T}^{\textsc{t}}$ refers to the transposed table
\STATE \quad \textsc{C}$'\leftarrow$ \textsc{C}$_1 \cup$ \textsc{C}$_2$
\STATE \quad \textsc{T}\textsubscript{C}$\leftarrow$ \textsc{T}{\fontfamily{qcr}\selectfont.filter}(C$'$) \\
$\triangleright$ Filter table \textsc{T} with columns \textsc{C}$'$ to get T$_\textsc{c}$\\
\underline{Row Extraction}
\STATE \quad \textsc{R}\textsubscript{1}$\leftarrow$ {\fontfamily{qcr}\selectfont row\textsubscript{sql}}(\textsc{T\textsubscript{c}}, \textsc{Q})$ \hfill \triangleright$ \textcolor{black}{SQL approach}}
\STATE \quad \textsc{R}\textsubscript{2}$\leftarrow$ {\fontfamily{qcr}\selectfont row\textsubscript{text}}(\textsc{T}\textsubscript{C}, \textsc{Q}, \textsc{R}\textsubscript{1})$ \hfill \triangleright$ \textcolor{black}{Text approach}}
\STATE \quad \textsc{R}$' \leftarrow$ \textsc{R}\textsubscript{1} $\cup$ \textsc{R}\textsubscript{2}
\STATE \quad \textsc{T}\textsubscript{CR} $\leftarrow$ \textsc{T}\textsubscript{C}{\fontfamily{qcr}\selectfont .filter}(R$'$) \\
$\triangleright$ Filter table $\textsc{T}_\textsc{c}$ with rows $\textsc{R}'$ to get $\textsc{T}_\textsc{cr}$\\
\textbf{Stage B: Adaptive Reasoning}\hfill
\IF {{\fontfamily{qcr}\selectfont math}(\textsc{Q}) $==$ {\fontfamily{qcr}\selectfont True}}
\STATE \textsc{E}\textsubscript{v} $\leftarrow$ {\fontfamily{qcr}\selectfont f\textsubscript{sql}}(\textsc{T}\textsubscript{\textsc{cr}}, \textsc{Q})
\STATE $\hat{A} \leftarrow$ {\fontfamily{qcr}\selectfont f\textsubscript{text}}(\textsc{T}\textsubscript{\textsc{cr}}, \textsc{Q}, \textsc{E}\textsubscript{v})\\
$\triangleright$ Use SQL generated \textsc{E}\textsubscript{v} with \textsc{T}\textsubscript{\textsc{cr}} to generate $\hat{A}$.
\ELSE
\STATE $\hat{A} \leftarrow$ {\fontfamily{qcr}\selectfont f\textsubscript{text}}(\textsc{T}\textsubscript{\textsc{cr}}, \textsc{Q}) \\ 
$\triangleright$ Use only the table $\textsc{T}_{\textsc{cr}}$ to generate $\hat{A}$
\ENDIF
\RETURN $\hat{A}$
\end{algorithmic}
\setlength{\intextsep}{0pt}
\end{algorithm}

\vspace{-0.25em}
\paragraph{Row Extraction.}
After obtaining the filtered table 
{\fontfamily{qcr}\selectfont T\textsubscript{C}} with relevant columns (Algorithm \ref{alg:cap}, line 4), \AlgName proceeds to the row extraction phase culminating in the query-specific table {\fontfamily{qcr}\selectfont T\textsubscript{CR}} (lines 5-8). The filtered table {\fontfamily{qcr}\selectfont T\textsubscript{C}} has fewer tokens than {\fontfamily{qcr}\selectfont T}, fitting within the token limit and enabling more efficient row extraction. As shown in Figure \ref{fig:main_pipeline}, the SQL query generated by {\fontfamily{qcr}\selectfont row\textsubscript{sql}} on the column-filtered table {\fontfamily{qcr}\selectfont T\textsubscript{C}} extracts the relevant rows {\fontfamily{qcr}\selectfont R\textsubscript{1}} ({\it `row 18'}). Although it returns the appropriate row, it is not sufficient to answer the question. To overcome these limitations, we use text-based row verification {\fontfamily{qcr}\selectfont row\textsubscript{text}} to obtain the missing rows, ultimately retrieving `{\it row 1}' and `{\it row 18}' ({\fontfamily{qcr}\selectfont R\textsubscript{2}}). Combining {\fontfamily{qcr}\selectfont R\textsubscript{1}} and {\fontfamily{qcr}\selectfont R\textsubscript{2}} gives the relevant rows {\fontfamily{qcr}\selectfont R'} ({\it `row 1', `row 18'}), resulting in the final table {\fontfamily{qcr}\selectfont T\textsubscript{CR}}.

\subsection{Adaptive Reasoning}
\textcolor{black}{We propose an adaptive reasoning framework that harnesses the strengths of textual reasoning while mitigating its quantitative limitations through symbolic reasoning. First, we prompt LLMs to leverage their language understanding capabilities to evaluate the query requirements. We apply symbolic reasoning for queries necessitating quantitative analysis and use the generated output ({\fontfamily{qcr}\selectfont E\textsubscript{v}}) along with the table ({\fontfamily{qcr}\selectfont T\textsubscript{CR}}) as the input to the final textual reasoning step.} For instance, in Figure \ref{fig:main_pipeline}, when faced with a question like \emph{`How long did it take the New York Americans to win the National Cup after 1936?'}, LLM needs to calculate the answer, and our pipeline prioritizes symbolic reasoning ({\fontfamily{qcr}\selectfont f\textsubscript{sql}}) through SQL-generated code. This code is executed on a SQL engine, enabling precise answers from the table {\fontfamily{qcr}\selectfont T\textsubscript{CR}}. The output from this SQL-based evaluation {\fontfamily{qcr}\selectfont E\textsubscript{v}} serves as additional evidence, supporting the final reasoning step ({\fontfamily{qcr}\selectfont f\textsubscript{text}}). By explicitly integrating this quantitative approach, \textit{our algorithm effectively addresses LLMs' limitations in handling quantitative reasoning}, thereby improving the accuracy and robustness when tackling questions that demand numerical reasoning.
\section{Experiments}

\paragraph{Benchmark Datasets.} We evaluate our method on three datasets covering fact verification, short-form, and long-form question-answering tasks using in-context examples. (a) TabFact \cite{chen2019tabfact}: A fact verification benchmark utilizing Wikipedia tables. We evaluated the test-small set, containing 2,024 statements and 298 tables, with each statement labeled Entailed ("True") or Refuted ("False"). (b) WikiTQ \cite{pasupat2015compositional}: The WikiTableQuestions (WikiTQ) dataset involves question-answering tasks over semi-structured Wikipedia tables. It includes a standard test set with 4,344 table-question pairs. (c) FeTaQA \cite{nan2022fetaqa}: FeTaQA (Free-Form Table Question Answering) comprises free-form questions requiring synthesis from multiple table sections for responses. It demands advanced reasoning and profound tabular data understanding, evaluated on a test set of 2003 samples.
\vspace{.25em}
\paragraph{Evaluation Metrics.} We tailor our evaluation metrics based on the task and dataset characteristics. For fact verification data sets such as TabFact, we evaluated performance using \emph{binary classification accuracy}. For short-form question-answering datasets such as WikiTQ, we assess accuracy by measuring the \emph{exact match} between predicted outputs and the gold-standard answers. For more complex tasks in FeTaQA, which involve long-form question answering, we evaluate performance using \emph{ROUGE-1}, \emph{ROUGE-2}, and \emph{ROUGE-L} metrics \cite{lin2004rouge}, comparing predicted outputs with long-form answers.
\vspace{.25em}
\paragraph{LLM Models.} In our research, we use state-of-the-art large language models (LLM) such as Gemini-1.5-Flash \cite{reid2024gemini}, PaLM-2 \cite{anil2023palm}, GPT-3.5-Turbo, GPT-4o-mini \cite{openai2023gpt}, and the open-source Llama-3-70B \cite{dubey2024llama} for table reasoning tasks. Our model inputs include in-context examples, the table, and the question for each step of the pipeline. All baselines with PALM-2 are from the propriety PALM-2 model, which is different from the public PALM-2 API (used in \AlgName and TabSQLify).

\paragraph{Baseline Methods.} We compare our method with (a) generic reasoning based on language models, and (b) table-manipulation reasoning based on language models.

\vspace{0.5em}
\textbf{(a) Generic Reasoning.} These methods direct the LLM to carry out the required downstream task based on the information from the table and the input question. They include End-to-End QA, which offers only the table and the question, and Few-Shot QA, which involves a few examples with the table-question-answer triplet alongside the table and the question. Chain-of-Thought prompts the LLMs to provide a supporting reasoning chain that leads to the answer. \textbf{(b) Table Manipulation.} These techniques involve several steps, with the initial stage dedicated to automatically pre-processing the table for the reasoning task. BINDER utilizes in-context examples to produce SQL or Python code containing LLM API calls to generate answers. DATER directs the language model to break down tables and questions, applying reasoning to the decomposed tables based on sub-questions. Chain-of-Table employs the table as an intermediate output in its reasoning process, iterating through tasks until the final answer is obtained. TabSQLify uses SQL to trim the table and then reasons based on the reduced table.
\subsection{Main Results}

Table \ref{table:performance} compares the performance of different methods on the TabFact, and WikiTQ datasets, across  GPT-3.5-Turbo and PaLM-2. This comparison involves evaluating against generic reasoning, table manipulation techniques, and \AlgName. Appendix \ref{sec:fetaqa} offers a comprehensive analysis of the FeTaQA dataset, including a comparison with baselines, human evaluation, and qualitative analysis.

\begin{table}[!h]
    \centering
    \small
    \vspace{-0.5em}
    \setlength{\tabcolsep}{3.5pt}
    \begin{tabular}{lcccc}
    \toprule
    &\multicolumn{2}{c}{\textbf{GPT-3.5-Turbo}} &\multicolumn{2}{c}{\textbf{PaLM-2}} \\
    \midrule
     & \textbf{TabFact} & \textbf{WikiTQ} & \textbf{TabFact} & \textbf{WikiTQ} \\
    \midrule
    \multicolumn{5}{l}{\textcolor{black}{\it Generic Reasoning}} 
    \\
    End-to-End QA & 70.45 & 51.84 & 77.92 & 60.59\\
    Few-shot QA & 71.54 & 52.56 & 78.06 & 60.33 \\
    CoT & 65.37 & 53.48 & 79.05 & 60.43 \\
    \midrule
    \multicolumn{5}{l}{\textcolor{black}{\it{Table Manipulation}}} \\
    BINDER & 79.17 & 56.74 & 76.98 & 54.88 \\
    DATER & 78.01 & 52.90 & 84.63 & 61.48\\
    Chain-of-Table* & 80.20 & 59.94 & \textcolor{black}{\textbf{86.61}} & 67.31 \\
    TabSQLify & 79.50 & 64.70 & 79.78 & 55.78 \\
    \midrule
    \textbf{\AlgName} & \textcolor{black}{\textbf{85.03}} & \textcolor{black}{\textbf{69.56}} & \textbf{86.51} & \textcolor{black}{\textbf{68.62}}\\ 
    \bottomrule
    \end{tabular}
\caption{\small \textbf{Comparison of various methods across datasets}. The results are reported from the Chain-of-Table paper for fair comparison. * Chain-of-Table uses a proprietary PaLM-2 model that is better than the publicly available version.}
\label{table:performance}
\end{table}

\textit{Analysis.} (a) On WikiTQ dataset, \AlgName surpasses all baselines with both GPT-3.5-Turbo and PaLM-2 models. \AlgName achieves an accuracy of 69.56\% on WikiTQ with GPT-3.5-Turbo, marking an improvement of 17.72\% over the vanilla GPT-3.5-Turbo model. (b) On the TabFact dataset, \AlgName outperforms all methods with GPT-3.5-Turbo, attaining an accuracy of 85.03\%, which is a 14.58\% improvement over the vanilla model. With PaLM-2, \AlgName achieves comparable performance with Chain-of-Table on TabFact while improving over it by 1.3\% on WikiTQ.

\paragraph{Comparison Across Methods.} Table \ref{table:extra_methods} compares \AlgName with ReAcTable \cite{zhang2023reactable} and SYNTQA \cite{zhang2024syntqa}, which use both textual and symbolic approaches for table reasoning. SYNTQA (GPT) uses the GPT model to decide between semantic and symbolic approaches, applying the chosen method for reasoning based on its selection. Additionally, it is compared with ALTER \cite{zhang2024alter}, which augments both queries and table data to facilitate reasoning and E5 \cite{zhang2024e5}, leveraging code generation capabilities of LLMs for table extraction followed by reasoning.

\begin{table}[!htbp]
    \centering
    \small
    \vspace{-0em}
    \scalebox{1}{
    \begin{tabular}{lcc}
    \toprule
     & \textbf{TabFact} & \textbf{WikiTQ} \\
    \midrule
    ReAcTable & 73.1 & 52.5 \\
    E5 & 75.6 & 50.9 \\
    ALTER & 84.3 & 67.4 \\
    SYNTQA (GPT) & - & 65.2 \\
    \midrule
    \textbf{\AlgName} & \textcolor{black}{\textbf{85.0}} & \textcolor{black}{\textbf{69.6}} \\ 
    \bottomrule
    \end{tabular}}
\caption{\small Comparison of various methods across datasets on GPT-3.5-Turbo.}
\label{table:extra_methods}
\end{table}

\emph{Analysis.} \AlgName outperforms methods that use symbolic and textual approaches, such as ReAcTable, SYNTQA(GPT), E5, and ALTER, showing the effectiveness of its complementary reasoning in a two-stage process. Appendix \ref{sec:methods} presents a comprehensive comparison of methods, including pre-trained models, fine-tuned architectures, and state-of-the-art LLM-based approaches. Our evaluation demonstrates that \AlgName consistently achieves superior performance across all benchmark metrics compared to all the baselines.


\paragraph{Performance Across LLMs.}
Table \ref{table:extra_models} compares the performance of more advanced models like Gemini-1.5-Flash, GPT-4o-mini, and the open-source Llama-3-70B on TabFact and WikiTQ datasets. The results emphasize \AlgName's generalizability across different models and consistent improvements across datasets. Even advanced models show significant benefits by using \AlgName.

\begin{table}[!htbp]
    \centering
    \small
    \setlength{\tabcolsep}{2pt}
    \begin{tabular}{lcccccc}
    \toprule
    &\multicolumn{2}{c}{\textbf{GPT-4o-mini}} &\multicolumn{2}{c}{\textbf{Gemini-1.5}} &\multicolumn{2}{c}{\textbf{Llama-3}} \\
    \midrule
     & \textbf{TF} & \textbf{WTQ} & \textbf{TF} & \textbf{WTQ} & \textbf{TF} & \textbf{WTQ} \\
    \midrule
    \multicolumn{7}{l}{\emph{Generic Reasoning}}\\
    End-to-End QA & 73.22 & 59.43 & 81.12 & 58.47 & 78.41 & 57.89\\
    CoT & 75.99 & 64.31 & 79.99 & 64.11 & 75.34 & 65.49\\
    \midrule
    \multicolumn{7}{l}{\emph{Table Manipulation}}\\
    TabSQLify & 78.30 & 68.74 & 79.50 & 63.92 & 60.70 & 66.85 \\
    Chain-of-Table & 85.09 & 68.53 & 86.95 & 70.05 & 85.86 & 70.76 \\
    \midrule
    \textbf{\AlgName} & \textcolor{black}{\textbf{89.42}} & \textcolor{black}{\textbf{74.93}} & \bf 89.08 & \bf 73.14 & \bf 89.23 & \bf 75.76\\ 
    \bottomrule
    \end{tabular}
\caption{\small Comparison of various models across datasets. {\bf TF}: TabFact; {\bf WTQ}: WikiTQ}
\label{table:extra_models}
\end{table}

\emph{Analysis.} (a) \AlgName shows much higher accuracy for GPT-4o-mini, scoring 89.42\% on TabFact and 74.93\% on WikiTQ, compared to the End-to-End QA and Chain of Thought methods, which achieve 73.22\% and 75.99\% on TabFact and 59.43\% and 64.31\% on WikiTQ. This marks a 16.2\% improvement on TabFact and 15.5\% on WikiTQ. (b) For Gemini-1.5-Flash, accuracy rises from 81.12\% and 58.47\% respectively to 89.08\% on TabFact and 73.14\% on WikiTQ. (c) Similarly, for Llama-3-70B, \AlgName improves the performance across both the datasets by a margin of 11\% on TabFact and 18\% on WikiTQ, showing consistent gains across models and datasets.
\subsection{Efficiency Analysis} 
\begin{figure}[!htbp]
\small
\centering
  \includegraphics[width=0.99\linewidth]{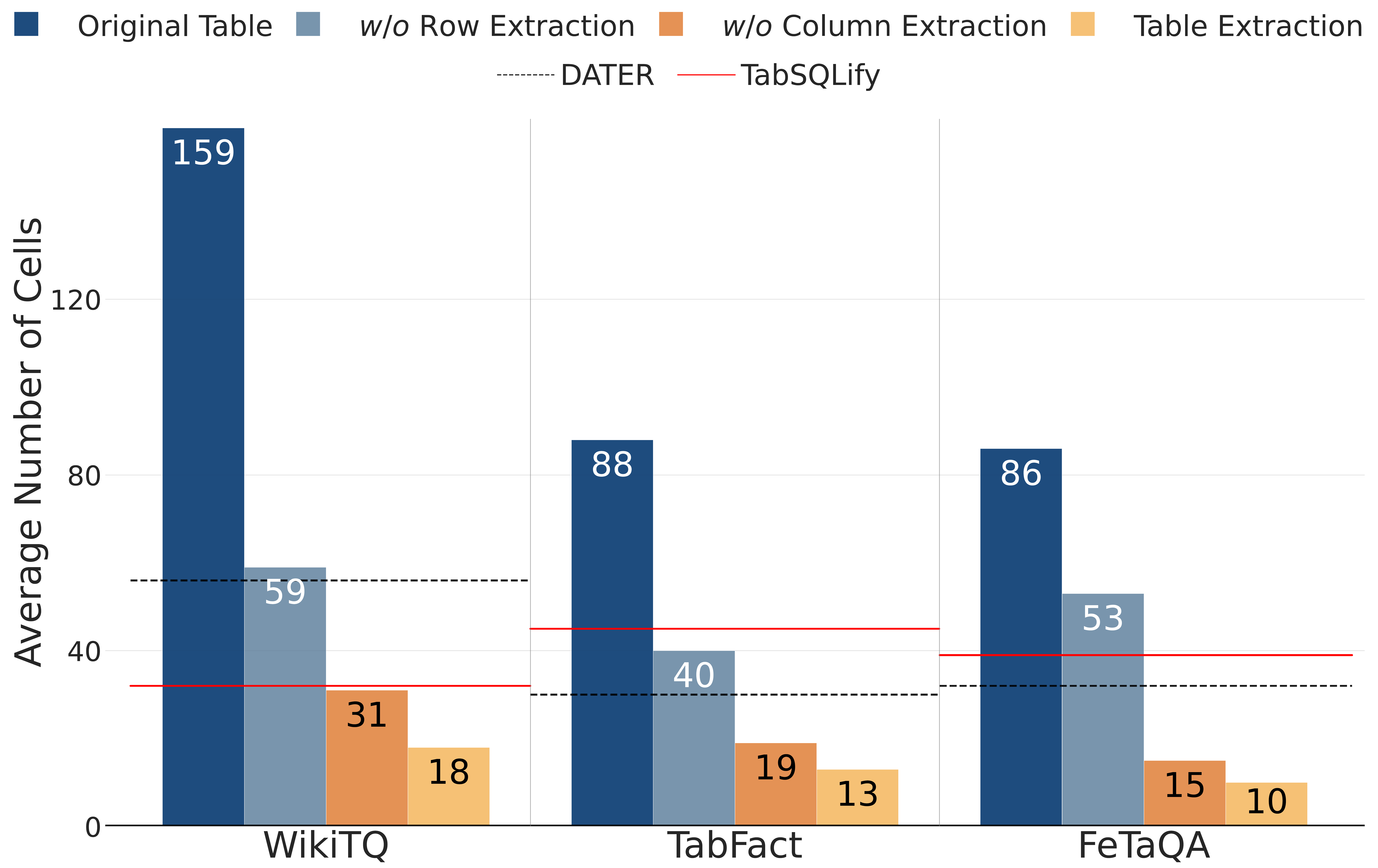}
  \caption{\small Comparison of average table cells in the final table.}
  \label{fig:reduction}
\end{figure}

\paragraph{Efficiency of Table Extraction.} Stage-1 in \AlgName involves the extraction of the table most relevant to the question. Our table extraction method uses a two-step chain to select the relevant columns followed by the rows. Figure \ref{fig:reduction} compares the number of average table cells for the extracted table for \AlgName, \AlgName without row extraction, and \AlgName without column extraction with other baselines.

\textit{Analysis.} \AlgName employs a structured, step-by-step process: it first extracts query-relevant columns before selecting relevant rows. This progressive filtering minimizes token usage at each stage, thereby reducing computational overhead. As a result, \AlgName significantly reduces the average number of processed table cells—dropping from 159 to 18 on the WikiTQ dataset, from 88 to 13 on TabFact, and from 86 to 10 on FeTaQA. In contrast, alternative table manipulation methods, such as TabSQLify and DATER, yield considerably higher values. By effectively filtering out irrelevant information, \AlgName enables LLMs to focus on the \emph{right evidence for right reasoning}, demonstrating superior efficiency across all datasets.


\paragraph{Efficiency on Longer Tables.} Table-CoT \cite{chen2023large} highlights that longer table size, with tables exceeding 30 rows, is a significant cause of erroneous generations. This observation is supported by a decline in LLM performance as the number of tokens in the table increases. In Table \ref{table:prompting_performance}, we classify LLM performance based on total tokens into three groups - small (< 2000 tokens), medium (2000 to 4000 tokens), and large (> 4000 tokens) and compare \AlgName with table manipulation like methods BINDER, DATER, Chain-of-Table, and TabSQLify.

\begin{table}[!htbp]
    \centering
    \small
    \setlength{\tabcolsep}{5pt}
    \begin{tabular}{lccc}
    \toprule
    \textbf{Method} & \textbf{Small} & \textbf{Medium} & \textbf{Large} \\
    \midrule
    BINDER & 56.54 & 25.13 & 6.41 \\
    DATER & 62.50 & 42.34 & 34.62 \\
    Chain-of-Table & 68.13 & 52.25 & 44.87 \\
    TabSQLify & 68.15 & 57.91 & 52.34 \\
    \midrule
    {\bf \AlgName} & \textcolor{black}{\textbf{71.64}} & \textcolor{black}{\textbf{65.20}} & \textcolor{black}{\textbf{64.84}} \\
    \bottomrule
    \end{tabular}
\caption{\small Performance of various methods on different table sizes on WikiTQ.}
\label{table:prompting_performance}
\end{table}

\textit{Analysis.} The results presented in Table \ref{table:prompting_performance} highlight the challenges LLMs face when reasoning over longer tables. Our findings confirm previous research, showing a significant decline in LLM performance as table sizes increase. While other table manipulation methods suffer from this scaling issue, \AlgName maintains consistent performance across all table sizes. This success is attributed to \AlgName’s efficient table extraction process, which reduces the table size by removing irrelevant data that would otherwise act as noise and hinder reasoning.


\paragraph{Resource Efficiency.}
\label{sec:symbolic_semantic}
In Table \ref{tab:generated_samples}, we analyze \AlgName by examining the total number of samples generated by LLMs. BINDER and DATER employ self-consistency techniques to refine their results, while Chain-of-Table follows an iterative sampling process. Specifically, BINDER generates 50 Neural-SQL samples using self-consistency, whereas DATER applies self-consistency at each step, producing 100 samples. In contrast, Chain-of-Table adopts a more resource-efficient approach, generating 25 samples across three steps: `Dynamic Plan', `Generate Args', and `Query'. TabSQLify generates the fewest samples, with a single generation for both table decomposition and query steps. However, relying on single outputs can reduce accuracy when the model fails to return a valid answer. To mitigate this, \AlgName integrates self-consistency by generating two outputs per stage, ensuring output validity. Each step, except `Query’, involves two generations for both SQL and text, totaling four steps for column and row retrieval. These additional steps act as safeguards against unreliable LLM outputs. Overall, \AlgName maintains efficiency, requiring only 6–10 sample generations.

\begin{table}[!htbp]
\footnotesize
    \centering
    \setlength{\tabcolsep}{4pt}
    \begin{tabular}{llc}
        \toprule
        \multirow{1}{*}{\textbf{Method}} & \textbf{\# samples / step} & \textbf{Total \# samples} \\
        \midrule
        {BINDER} & Neural SQL: 50 & 50 \\ \midrule
        \multirow{4}{*}{{DATER}} & Decompose Table: 40 & \multirow{4}{*}{100}\\
        & Generate Cloze: 20 & \\
        & Generate SQL: 20 & \\
        & Query: 20 & \\ \midrule 
        \multirow{3}{*}{{Chain-of-Table}} & Dynamic Plan $\leq$ 5 & \multirow{3}{*}{$\leq$ 25}\\
        & Generate Args $\leq$ 19 & \\
        & Query: 1 &  \\ \midrule
        \multirow{2}{*}{{TabSQLify}} & Table Decompose: 1 & \multirow{2}{*}{2} \\
        & Query: 1  & \\ \midrule
        \multirow{3}{*}{\textbf{\AlgName}} & Column Extraction: 2-4 & \multirow{3}{*}{\text{6-10}}\\ 
        & Row Extraction: 2-4 & \\
        & Query: 2 & \\ \bottomrule
    \end{tabular}
    \caption{\small Number of generated samples for different methods.}
    \label{tab:generated_samples}
\end{table}

\subsection{Error Analysis}
\begin{figure}[!htbp]
\centering
  \includegraphics[width=\linewidth]{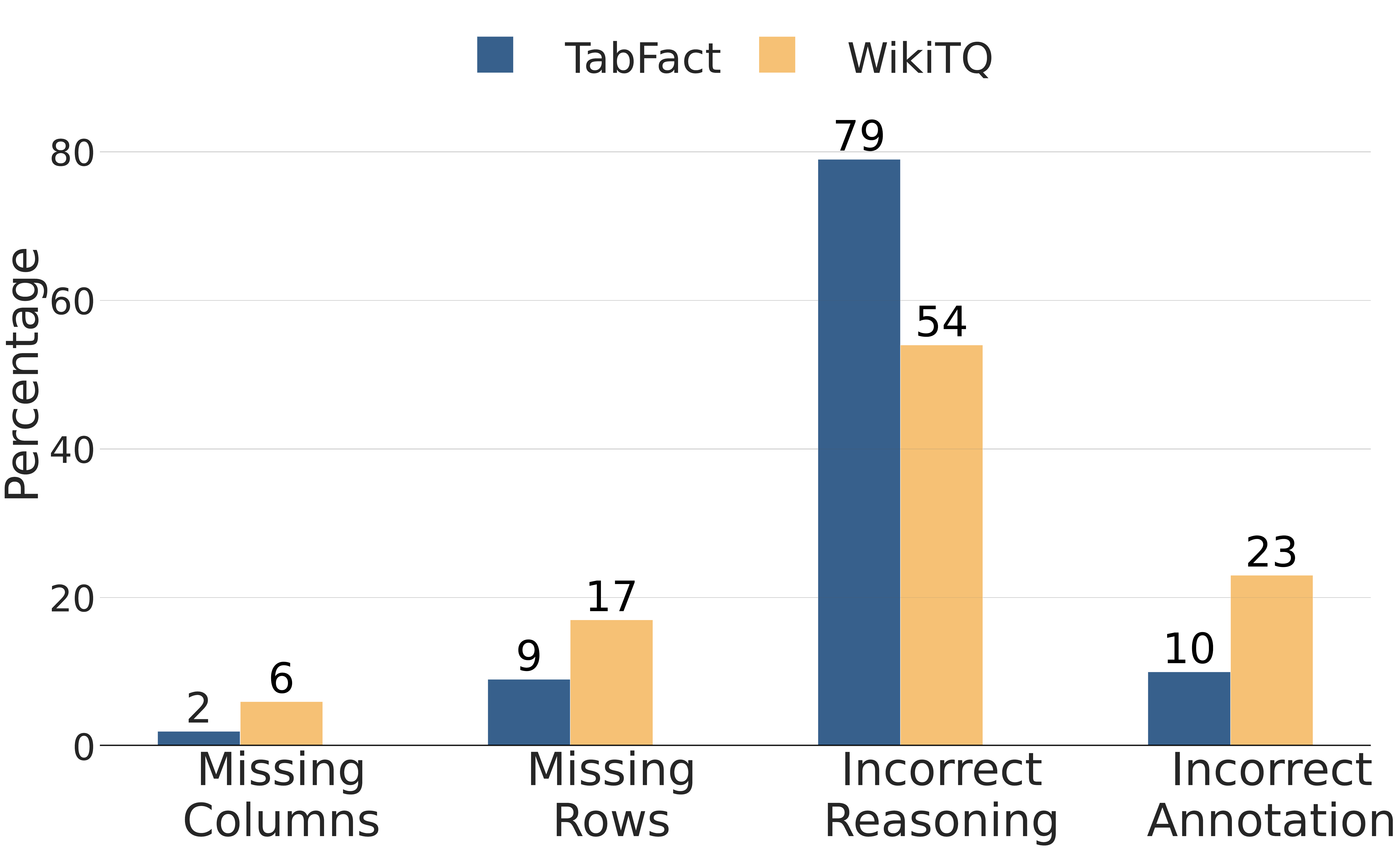} 
  \caption{\small Error distribution on 100 error samples across datasets for \AlgName (GPT-3.5-Turbo).} 
  \label{fig:errors}
\end{figure}

\paragraph{\AlgName.} The disjoint, step-wise nature of \AlgName enables identifying and analyzing failures. Figure \ref{fig:errors} shows 100 randomly selected instances from the Tabfact and WikiTQ datasets where \AlgName returns incorrect answers. In this study, `Missing Columns' and `Missing Rows' refer to missing necessary columns and rows, respectively. `Incorrect Reasoning' occurs when \AlgName extracts the correct table, but LLM fails to produce the correct answer. `Incorrect Annotations' include semantically identical answers in different formats, ambiguous questions, and incorrect gold answers. Figure \ref{fig:errors} shows that for 100 TabFact failures, 2\% are missing columns, 9\% missing rows, 79\% incorrect LLM reasoning, and 10\% incorrect annotations. For WikiTQ, 6\% are missing columns, 17\% missing rows, 54\% incorrect reasoning, and 23\% incorrect annotations.

\textit{Analysis.} Fewer errors in column and row extraction indicate that \AlgName effectively retrieves the necessary table data.  This improvement allows us to "shift left" by addressing issues earlier in the pipeline, enhancing overall performance. The higher percentage of errors in reasoning does not reflect poorly on the LLM; instead, it underscores the effectiveness of our table extraction process.

\begin{figure}[!htbp]
\centering
  \includegraphics[width=\linewidth]{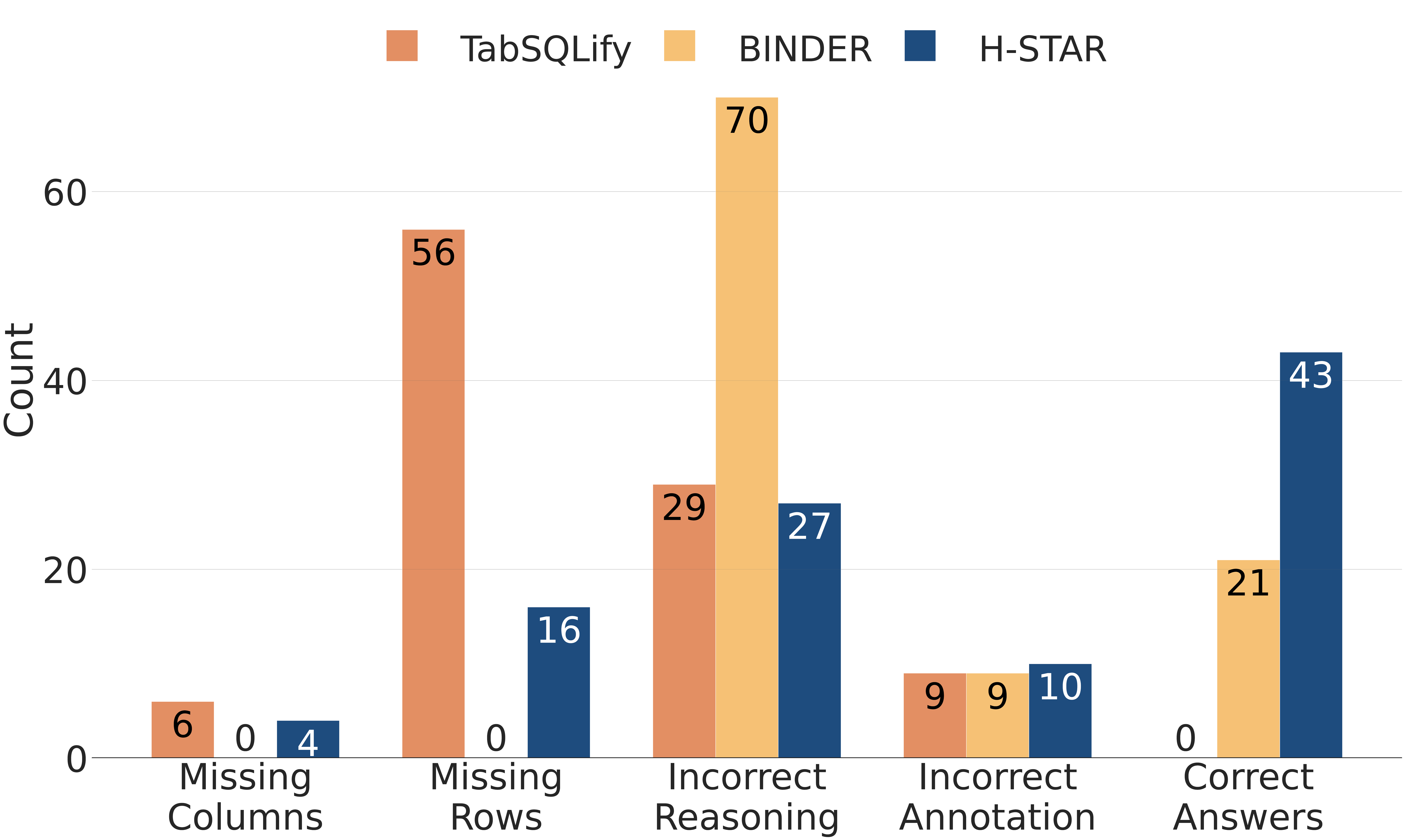} 
  \caption{\small  Analysis of error types in 100 samples from WikiTQ where TabSQLify fails. \textbf{Note}: As we move from left to right, the total samples decrease for each stage in the pipeline.} 
  \label{fig:error_comp}
\end{figure}

\paragraph{\AlgName vs Others.}
Figure \ref{fig:error_comp} compares errors in \AlgName, with TabSQLify, and BINDER on 100 WikiTQ samples where TabSQLify fails. TabSQLify extracts wrong tables for 62 out of 100 (6 missing columns, 56 missing rows). Out of the 38 samples that remain, it incorrectly reasons on 29 of the samples (76\%). 
BINDER does not handle table extraction and instead relies on generating multiple neural SQL queries for final reasoning. However, it fails to reason correctly in 70 out of 100 instances.

\emph{Analysis.} In contrast, \AlgName proves more effective in both table extraction and reasoning, generating correct answers on 43 samples. Additionally, it demonstrates fewer errors in table extraction (4 missed columns; 16 missed rows) and reasoning (27 of the remaining 80 samples, i.e. 34\%).
\section{Ablation Study}
To evaluate the significance of the two primary stages in \AlgName: (a) \textbf{Table extraction} and (b) \textbf{Adaptive reasoning}, we conducted an ablation study, removing one stage at a time. For the first condition, the final adaptive reasoning stage was substituted with a basic chain-of-thought reasoning process while maintaining the table extraction stage. Conversely, the second condition retained the adaptive reasoning stage but omitted the table extraction stage, specifically removing the row extraction and column extraction steps individually and then collectively. A detailed assessment of the contribution of each stage to the overall performance of the method is shown in Table \ref{table:ablation1}. It highlights the importance of each step in the pipeline. Removing any step can result in a performance loss especially adaptive reasoning which causes a 7\% drop for both TabFact and WikiTQ datasets. For a detailed error case study and further analysis, refer to the Appendix \ref{sec:row_col}.
\begin{table}[!htbp]
    \small
    \centering
    \scalebox{1}{
    \begin{tabular}{lcc}    
    \toprule
    \textbf{Method} & \textbf{TabFact} & \textbf{WikiTQ} \\
    \midrule
    \textbf{\AlgName} & \bf 86.51 & \textcolor{black} {\bf{68.62}} \\
    \midrule
    \textit{w/o} row extraction & \textcolor{black}{{86.17}} & 66.30 \\
    \textit{w/o} column extraction & 84.04 & 67.03 \\
    \textit{w/o} table extraction & 83.79 & 63.58 \\
    \textit{w/o} adaptive reasoning & 79.35 & 61.47 \\
    \bottomrule
    \end{tabular}}
\caption{\small Performance on TabFact and WikiTQ datasets.}
\label{table:ablation1}
\end{table}

\subsection{Impact of the Hybrid Approach}
We perform a quantitative study of the impact of the hybrid approach in \AlgName. Figures \ref{fig:column} and \ref{fig:row} (Appendix \ref{sec:row_col}) show how the use of only one of the approaches results in insufficient outputs for column and row extraction tasks, respectively. Furthermore, Table \ref{table:ablation2} shows that not employing a hybrid approach results in a significant performance loss. 

\begin{table}[!htbp]
\small
    \centering
    \setlength{\tabcolsep}{7pt}
    \begin{tabular}{lcc}
    \toprule
    \textbf{Method} & \textbf{TabFact} & \textbf{WikiTQ} \\
    \midrule
    CoT & 79.05 & 60.43 \\
    \AlgName$_{\text{text}}$ & 79.43 & 61.47 \\
    \midrule
    Text-to-SQL & 52.05 & 42.03 \\
    \AlgName$_{\text{sql}}$ & 58.50 & 46.09 \\
    \midrule
    \bf \AlgName & \textcolor{black}{\bf 86.51} & \textcolor{black}{\bf 68.62} \\
    \bottomrule
    \end{tabular}
\caption{\small {Performance using only text-based/SQL-based reasoning}; {\AlgName\textsubscript{text}}: \AlgName with textual reasoning only; { \AlgName\textsubscript{sql}}: \AlgName with symbolic reasoning only.}
\label{table:ablation2}
\end{table}

\emph{Analysis.} SQL methods though efficient with numerical reasoning, are highly sensitive to data variations such as irregular formatting and mixed data types, leading to a performance drop. In contrast, textual reasoning is more resilient, offering fuzzy matching and better interpretation of data. By combining textual and symbolic reasoning, \AlgName uses their complementary strengths, achieving superior performance. 

\subsection{Symbolic vs Semantic Approaches}
As shown in Table \ref{tab:ablation_2} removing SQL extraction drops the accuracy to 85.22\% on TabFact and 64.39\% on WikiTQ, and removing text extraction causes a drop to 83.74\% and 60.31\%, respectively. SQL reasoning removal results in 84.48\% and 64.76\%, but omitting text reasoning causes the largest drop, to 58.70\% and 54.35\%.
\begin{table}[!htbp]
\small
    \centering
    \setlength{\tabcolsep}{10pt}
    \begin{tabular}{lcc}
        \toprule
        \bf Method	& \bf TabFact & \bf WikiTQ \\
        \midrule
        \bf H-STAR	& \bf 86.51	& \bf 68.62 \\
        \midrule
        \textit{w/o} SQL extraction	& 85.22	& 64.39 \\
        \textit{w/o} text extraction & 83.74 & 60.31 \\
        \textit{w/o} SQL reasoning	& 84.48	& 64.76 \\
        \textit{w/o} text reasoning	& 58.70	& 54.35 \\
    \bottomrule
    \end{tabular}
    \caption{\small Performance of H-STAR after systematically removing symbolic part and semantic parts from: (1) table extraction (both row and column extraction); (2) adaptive reasoning.}
    \label{tab:ablation_2}
\end{table}

\textit{Analysis.} Table \ref{tab:ablation_2} shows that text-based approaches excel in both table extraction and reasoning. The evaluation datasets often contain noisy data that SQL-based approaches struggle with due to their reliance on structured schemas. Since text-based reasoning methods are more effective at handling such irregularities, they achieve reasonably high scores, albeit lower than our hybrid approach.
\section{Key Findings}

Firstly, our evaluation demonstrates that our hybrid approach achieves substantial improvements, surpassing the performance of previous state-of-the-art methods across various table reasoning tasks. Secondly, the quantitative analysis demonstrates that our `multi-view' approach extracts tables specific to the query. Furthermore, the qualitative analysis highlights fewer errors in our table extraction method compared to prior approaches, confirming a decrease in irrelevant information. 

Thirdly, our analysis indicates consistent performance even with longer tables, emphasizing our method's effectiveness in accurately extracting relevant information by filtering out noise. Lastly, the ablation study shows that decomposing the task into sub-tasks significantly enhances the overall performance, with each sub-task playing a crucial role in achieving superior results. Moreover, it also highlights the constraints of relying solely on either text or SQL approaches, which are effectively addressed by our \AlgName approach. Together, these findings emphasize the substantial advantages of our hybrid and multi-view approach in addressing complex table reasoning tasks.

\section{Related Work}
Table reasoning tasks require the ability to reason over unstructured queries and structured or semi-structured tables. Traditional approaches like TAPAS \cite{herzig2020tapas}, TAPEX \cite{liu2021tapex}, TABERT \cite{yin2020tabert}, TURL \cite{deng2022turl}, PASTA \cite{gu2022pasta}  work on pre-training language models jointly on large-scale tabular and text data to reason in an end-to-end manner. Advancements in LLMs, allow them to learn from in-context examples, reducing inference costs. Text-to-SQL \cite{rajkumar2022evaluating} and Program-of-Thought \cite{chen2023program} use symbolic methods to solve table-based tasks via Python/SQL programs. Textual-based reasoning techniques such as Table-CoT \cite{chen2023large} and Tab-CoT \cite{ziqi-lu-2023-tab} extend prompting methods such as zero- and few-shot CoT for tabular reasoning.

The decomposition of problems into smaller and manageable tasks has proven effective in solving complex reasoning challenges \cite{zhou2022least, khot2022decomposed}. Recent techniques in table reasoning follow this approach, either by breaking tasks into fixed sub-tasks \cite{cheng2022binding, ye2023large, nahid2024tabsqlify} or by employing iterative methods \cite{jiang2023structgpt, zhang2023reactable, wang2023chain}. BINDER \cite{cheng2022binding} is an SQL-based approach that modifies SQL statements to include LLM API calls within SQL statements. \textcolor{black}{ALTER \cite{zhang2024alter} augments both the queries along with the table data.} DATER \cite{ye2023large} and TabSQLify \cite{nahid2024tabsqlify} are table decomposition methods that use semantic and symbolic reasoning, respectively. Chain-of-Table \cite{wang2023chain} uses a textual reasoning approach to update tables iteratively before the final reasoning step. SYNTQA \cite{zhang2024syntqa} is an ensemble approach employing an answer selection mechanism that selects between text-to-SQL and text-based models. The works most closely related to our approach are ReAcTable \cite{zhang2023reactable} and E5 \cite{zhang2024e5}. ReAcTable extends the ReAct framework \cite{yao2023react} to table reasoning by employing step-by-step reasoning, where it iteratively generates sub-tables using external tools like SQL and Python. Similarly, E5 introduces a multi-step process for hierarchical table question answering, interpreting table structures, generating and executing code, and performing reasoning over the results to derive answers. In contrast, \AlgName employs a fixed two-stage pipeline for table extraction and reasoning, utilizing both symbolic and semantic approaches complementarily at each stage, resulting in improved performance. A comprehensive comparison of the methodological differences and their implications for performance is provided in Appendix \ref{sec:react}.




\section{Conclusion}
In this study, we present \AlgName, a novel method that effectively integrates semantic and symbolic approaches, demonstrating superior performance compared to existing methods in tasks involving table reasoning. Our approach involves a two-step LLM-driven process: firstly, employing `multi-view' table extraction to retrieve tables relevant to a query, and then, implementing adaptive reasoning to select the optimal reasoning strategy based on the input query. We address prior bottlenecks with efficient extraction and reasoning, leading to improved overall performance, particularly on longer tables. Our results highlight the need to move beyond relying solely on either technique and demonstrate the effectiveness of an integrated approach that combines the advantages of both methods. Future directions involve testing the adaptability of our methods to semi-structured, complex hierarchical, and relational tables. Enhancing the reasoning process through techniques such as self-consistency and self-verification shows promising potential.

\section*{Limitations}
Our current work has primarily focused on a subset of table reasoning tasks using datasets sourced from Wikipedia. While this has laid a solid foundation, it limits exploration into diverse reasoning tasks such as table manipulation, text-to-table generation, and table augmentation, which could provide valuable insights and enhance our approach's capabilities. Additionally, our method's generalizability is confined to Wikipedia-based datasets, restricting its application to other domains that require specific domain knowledge, which our current approach lacks. Extending our approach to different domains may necessitate integrating domain-specific knowledge to ensure effective reasoning.

Furthermore, our evaluation has been limited to relatively straightforward table structures. Handling more complex data representations such as semi-structured tables, hierarchical tables, and relational databases remains unexplored territory. These structures present unique challenges that our current approach may not effectively address. Finally, our study has focused solely on the English language, potentially limiting its applicability to languages with different linguistic complexities that we have not accounted for.

\section*{Ethics Statement}
We, the authors, affirm that our work adheres to the highest ethical standards in research and publication. We have carefully considered and addressed various ethical issues to ensure the responsible and fair use of computational linguistics methodologies. To facilitate reproducibility, we provide detailed information, including code, datasets (all publicly available and in compliance with their respective ethical standards), and other relevant resources. Our claims align with the experimental results, though some stochasticity is expected with black-box large language models, which we minimize by maintaining a fixed temperature. We provide comprehensive details on annotations, dataset splits, models used, and prompting methods, ensuring our work can be reliably reproduced.

\section*{Acknowledgments}
This research was partly sponsored by the U.S. National Science Foundation (NSF) under Grant No. 2416728, Army Research Office under Grant Number W911NF-20-1-0080, and ONR Contract N00014-23-1-2364. The views and conclusions contained in this document are those of the authors and should not be interpreted as representing the official policies, either expressed or implied, of the Army Research Office or the U.S. Government. The U.S. Government is authorized to reproduce and distribute reprints for Government purposes notwithstanding any copyright notation herein. We thank the authors of Tabsqlify and BINDER for sharing their code, and Yerram Varun for his help in setting up the implementation process. We also extend our gratitude to Jennifer Sheffield from UPenn for research support. Lastly, we appreciate the reviewing team’s insightful comments.


\bibliography{anthology,custom}
\bibliographystyle{acl_natbib}

\appendix

\section{Supplementary Analysis}
\label{sec:additional results}
\subsection{FeTaQA}
\label{sec:fetaqa}
The results for the FeTaQA dataset using the ROUGE scores \cite{lin2004rouge} are given in Table \ref{table:fetaqa}. We observe an incremental improvement in the scores when compared to other methods on the GPT-3.5-Turbo. \emph{The ROUGE scores focus on lexical similarities while ignoring the semantic similarity between predicted and gold outputs}. These metrics often fail to capture improvements with in-context learning, as the model doesn't learn the long-form text style from just an instruction or a few examples.

\begin{table}[!htbp]
    \centering
    \small
    \setlength{\tabcolsep}{2pt}
    \scalebox{0.9}{
    \begin{tabular}{lccc}
    \toprule
    \textbf{Prompting} & \textbf{ROUGE-1} & \textbf{ROUGE-2} & \textbf{ROUGE-L} \\
    \midrule
    \textbf{PaLM-2*} & & \\
    DATER & 0.63 & 0.41 & 0.53\\
    Chain-of-Table & 0.66 & 0.44 & 0.56\\
    \midrule
    \textbf{GPT-3.5-Turbo} & & \\
    Table-CoT & 0.62 & 0.39 & 0.51\\
    TabSQLify & 0.58 & 0.35 & 0.48\\
    \midrule
    \bf \AlgName & 0.62 & 0.39 & 0.52\\
    \bottomrule
    \end{tabular}}
\caption{\small Performance on FeTaQA dataset. PaLM-2* refers to the Google proprietary PaLM-2 model.}
\label{table:fetaqa}
\end{table}

\begin{figure}[htbp]
    \centering
    \includegraphics[width=0.85\columnwidth]{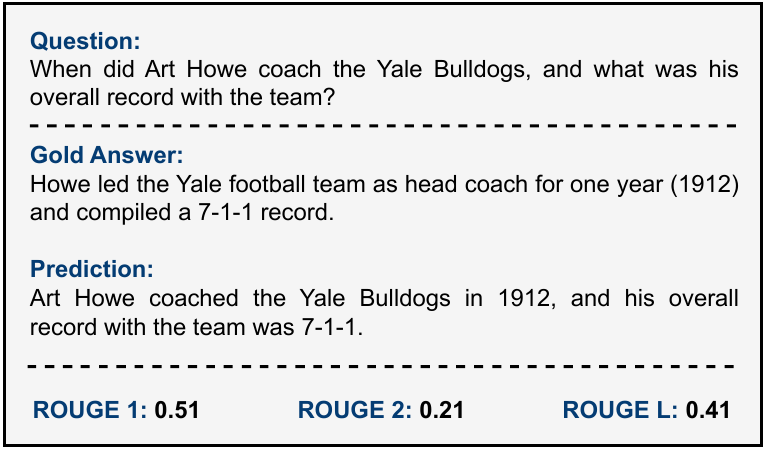}
    \caption{\small An example of FeTaQA representing low ROUGE scores for correct answers.}
    \label{fig:fetaqa_ex}
\end{figure}

Figure \ref{fig:fetaqa_ex} illustrates an example where the generated output, despite being correct in answering the question, is penalized by the \emph{ROUGE} metric. This highlights the limitations of the metric in evaluating the correctness of the generated responses.

\paragraph{Qualitative Analysis.}
In this section, we perform a qualitative analysis on 100 samples of FeTaQA with TabSQLify \cite{nahid2024tabsqlify}. The study aims to compare the generations of \AlgName with TabSQLify by evaluating the answer quality. The answers are evaluated based on their fluency and adequacy compared to the gold answers. Table \ref{tab:fetaqa_comp} shows that \AlgName has better quality generations in 26\% samples, as opposed to only 18\% for TabSQLify. `Both' indicates samples where both the algorithms perform on par. This can be attributed to the use of the same LLM. Many failures for TabSQLify are derived from an insufficient table, further highlighting the limitations of using a single-view approach for table extraction. 

\begin{table}[!htbp]
\small
    \centering
    \setlength{\tabcolsep}{10pt}
    \centering
    \begin{tabular}{lc}
    \toprule
    \bf Method & \bf \% of samples \\
    \midrule
    Both & 56\%\\
    TabSQLify & 18\%\\
    \midrule
    \bf \AlgName & 26\%\\
    \bottomrule
    \end{tabular}
    \caption{\small Comparison of \AlgName with TabSQLify on 100 samples.}
    \vspace{-0.75em}
    \label{tab:fetaqa_comp}
\end{table}

\begin{figure}[!htbp]
    \centering
    \vspace{-0.5em}
    \includegraphics[width=0.88\columnwidth]{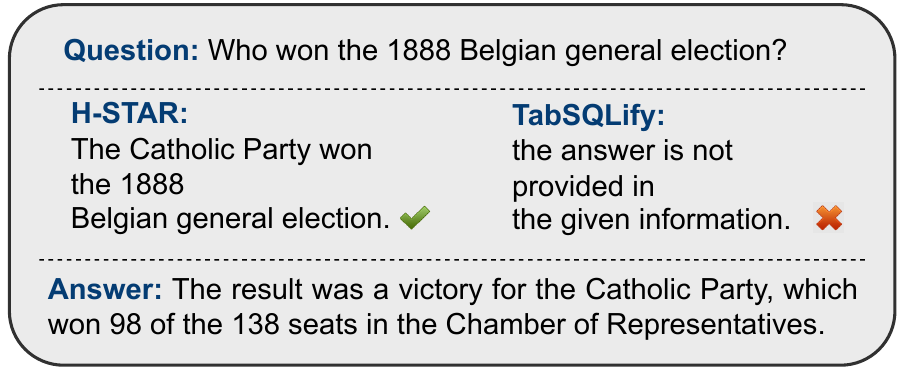}
    \caption{\small An example of FeTaQA dataset where TabSQLify fails as a result of insufficient table evidence.}
    \label{fig:fetaqa_fail}
   \vspace{-0.5em}
\end{figure}

\begin{figure}[!htbp]
    \centering
    \includegraphics[width = 0.88\linewidth]{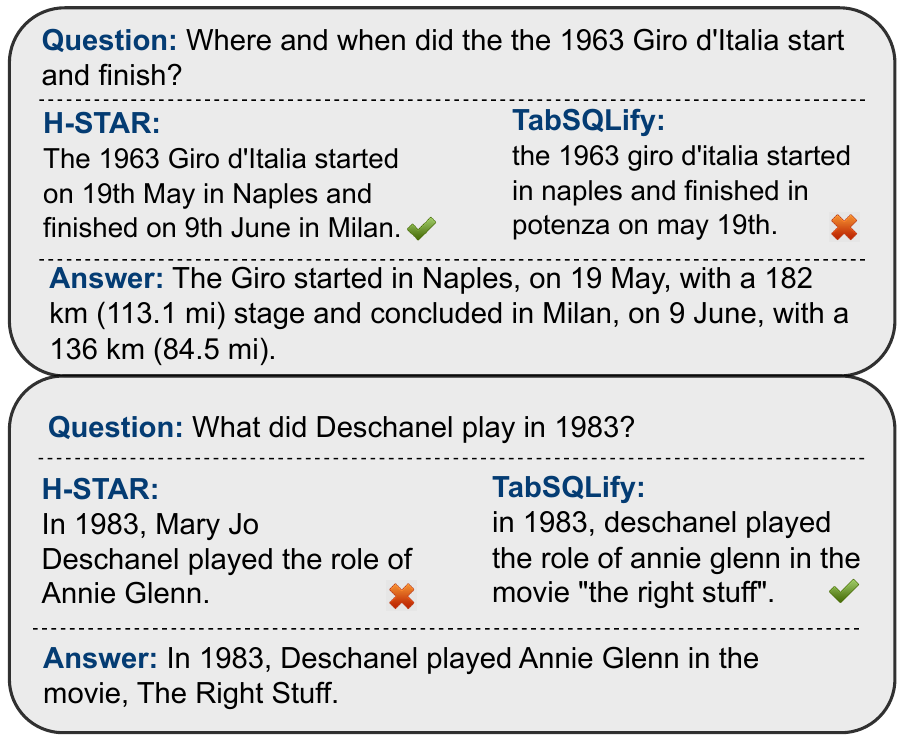}
    \caption{\small Comparison of the outputs with TabSQLify.}
    \label{fig:fetaqa_comp}
    \vspace{-0.5em}
\end{figure}

Figure \ref{fig:fetaqa_fail} presents an example in which TabSQLify returns \emph{`information not provided'} due to an incorrect table. It can be seen that \AlgName generates an answer comparable with the ground truth for the same. Figure \ref{fig:fetaqa_comp} visually depicts the outputs from both methods. It illustrates the evaluation process undertaken to compare the output quality. 

\paragraph{Human Evaluation.} We further evaluate the quality of our generations using human evaluation. We refer to \cite{nan2022fetaqa} and evaluate our generations on four factors: {\bf (1) Fluency} based on the nature and grammar of the output;  {\bf (2) Correctness} measuring the degree of correctness of the prediction; {\bf (3) Adequacy} to measure whether all the necessary information is contained in the output; {\bf (4) Faithfulness} to check whether the generation is grounded in the final extracted table used for reasoning. We ask five internal human annotators to assign a score of 1-5 for each criterion and report the percentage of samples having values of 4 or 5. Table \ref{table:human_eval} shows the results by five annotators for our method and compares it with other prior methods. The performance pattern of \AlgName is similar to human performance with comparable fluency and faithfulness while matching the trend of having higher adequacy over correctness. Although we marginally outperform other methods, there is still scope for improvement.

\begin{table}[!htbp]
    \centering
    \small
    \setlength{\tabcolsep}{3pt}
    \scalebox{0.9}{
    \begin{tabular}{lcccc}
    \toprule
    \textbf{Method} & \textbf{Fluent} & \textbf{Correct} & \textbf{Adequate} & \textbf{Faithful} \\
    \midrule
    {T5-large} & 94.6 & 54.8 & 50.4 & 50.4 \\
    {Human} & 95 & 92.4 & 95.6 & 95.6 \\
    \midrule
    {TableCoT} & 96 & 82 & 75 & 87 \\
    {TabSQLify} & 97 & 88 & 84 & 93 \\
    \midrule
    \textbf{\AlgName} & 96.6 & 87.6 & 89.6 & 94 \\
    \bottomrule
    \end{tabular}}
\caption{\small Human evaluation results on FeTaQA dataset.}
\label{table:human_eval}
\end{table}

\subsection{Comparison With Diverse Methods}
\label{sec:methods}
Tables \ref{tab:wikitq} and \ref{tab:tabfact} show results of other baseline methods on WikiTQ and TabFact datasets respectively. Our method, \AlgName, outperformed all other models by substantial margins. Despite using the smaller, less capable GPT-4o-mini, H-STAR outperforms E5 and ReAcTable which use GPT-4 on both TabFact and WikiTQ. Furthermore, H-STAR (GPT-3.5-Turbo) outperforms E5 (GPT-4) and ReAct (GPT-4) on WikiTQ.
\begin{table}[!htbp]
\small
    \centering
    \setlength{\tabcolsep}{4pt}
    \centering
    \scalebox{0.88}{
    \begin{tabular}{llc}
    \toprule
    \textbf{Model} & \textbf{Method} & \textbf{Accuracy} \\
    \midrule
   \multirow{5}{*}{Pre-trained/fine-tuned} & TaPas & 48.8 \\
    & GraPPa & 52.7 \\
    & LEVER & 62.9 \\
    & ITR & 63.4 \\
    & SYNTQA(GPT)* & 70.4 \\
    \midrule
    \multirow{5}{*}{Codex\tablefootnote{OpenAI Codex model is not available publically anymore.}} & GPT-3 CoT & 45.7 \\
    & TableCoT & 48.8 \\
    & DATER & 65.9 \\
    & BINDER & 61.9 \\
    & ReAcTable & 65.8 \\
    \midrule
    \multirow{4}{*}{GPT-3.5-Turbo} & ReAcTable & 52.5 \\
    & E5 & 50.9 \\
    &  TableCoT & 52.4 \\
    & StructGPT & 52.2 \\ 
    & ALTER & 67.4 \\ 
    & SYNTQA(GPT)* & 65.2 \\ \midrule
    \multirow{2}{*}{GPT-4} & ReAcTable & 57.3 \\
     & E5 & 65.5 \\ \midrule
    GPT-3.5-Turbo & \bf \multirow{2}{*}{\AlgName}  & \bf 69.6 \\
    GPT-4o-Mini &  & \bf 74.9 \\
    \bottomrule
    \end{tabular}}
    \caption{\small Comparison of \AlgName with additional methods on WikiTQ dataset. * SYNTQA (GPT) uses GPT-3.5-Turbo to choose between SQL/text reasoning and uses: (1) fine-tuned models for reasoning; (2) GPT-3.5-Turbo for reasoning.}
    \label{tab:wikitq}
\end{table}

\begin{table}[!htbp]
\small
    \centering
    \setlength{\tabcolsep}{4pt}
    \centering
    \scalebox{0.88}{
    \begin{tabular}{llc}
    \toprule
    \bf Model & \bf Method & \bf Accuracy \\
    \midrule
    \multirow{7}{*}{{Pre-trained/fine-tuned}} & Table-BERT & 68.1 \\
    & LogicFactChecker & 74.3 \\
    & SAT & 75.5 \\
    & TaPas & 83.9 \\
    & TAPEX & 85.9 \\
    & SaMoE & 86.7 \\
    & PASTA & 90.8 \\
    \midrule
    \multirow{4}{*}{{Codex}} & TableCoT & 72.6 \\
    & DATER & 85.6 \\
    & BINDER & 85.1 \\
    & ReAcTable & 83.1 \\
    \midrule
    \multirow{3}{*}{{GPT-3.5-Turbo}} & ReAcTable & 73.1 \\
    & E5 & 75.6 \\
    & TableCoT & 73.1 \\ 
    & ALTER & 84.3 \\ \midrule
    \multirow{2}{*}{GPT-4} & ReAcTable & 83.7 \\
     & E5 & 88.8 \\ \midrule
    GPT-3.5-Turbo & \bf \multirow{2}{*}{\AlgName} & \bf 85.0 \\
    GPT-4o-Mini &  & \bf 89.4 \\
    \bottomrule
    \end{tabular}}
    \caption{\small Comparison of \AlgName with additional methods on TabFact dataset.}
    \label{tab:tabfact}
\end{table}

\subsection{Advantages of using Transposed Table}
\label{sec:transposed}
Table \ref{tab:transposed_table} compares the performance of H-STAR using original versus transposed tables for text-based column verification. In TabFact, H-STAR achieves 86.51\% accuracy with transposed tables versus 85. 22\% with the original tables. On WikiTQ, accuracy is 68.62\% with transposed tables, compared to 66. 59\% with the original tables.
\begin{table}[!htbp]
\small
    \centering
    \setlength{\tabcolsep}{4pt}
    \begin{tabular}{lcc}
        \toprule
        \bf Method	& \bf TabFact	& \bf WikiTQ \\
        \midrule
        \bf H-STAR (with transposed table)	& 86.51	& 68.62 \\
        \bf H-STAR (with original table) & 85.22 & 66.59 \\
    \bottomrule
    \end{tabular}
    \caption{\small Comparison between H-STAR using (a) transposed table and (b) original table for text-based column extraction.}
    \label{tab:transposed_table}
\end{table}

\emph{Analysis.} Transposing tables leads to better performance than using the original table for column verification. The transposed table approach is more effective for complex datasets such as WikiTQ, which are harder to reason on and contain longer tables.

\section{Comparison with Other Methods}
\label{sec:react}

While H-STAR shares similarities with ReAct and E5 in combining textual reasoning with tool-based methods, there are key differences. Like E5, H-STAR employs table extraction to filter irrelevant data and leverages LLMs' code generation capabilities for table reasoning. However, there are significant differences.
\paragraph{1. Integration of Symbolic and Semantic Reasoning.}
H-STAR combines SQL-based and text-based reasoning, while ReAct merges textual reasoning with actions. While ReAct uses tools primarily for actions and E5 applies code generation only for table extraction, both rely solely on semantic approaches for reasoning. In contrast, H-STAR applies SQL for quantitative tasks and text for lexical queries.

\paragraph{2. Fixed Multi-Stage Approach.}
ReAct prompts LLMs to alternate between reasoning and actions, allowing interaction with external sources. H-STAR, on the other hand, splits tabular reasoning into two stages: (1) Table Extraction, and (2) Adaptive Reasoning, blending SQL and text reasoning. This hybrid approach sets H-STAR apart from ReAct’s focus on textual reasoning and interaction with the environment.
\paragraph{3. Table Extraction Approach.}
H-STAR distinguishes itself by combining symbolic and semantic methods for table extraction, rather than relying exclusively on either approach. Furthermore, 
H-STAR uses a `multi-view' approach for table extraction, identifying relevant columns from both original and transposed tables before selecting rows, reducing irrelevant data, and improving focus. In contrast, ReAct does not employ any extraction method.
\paragraph{4. Adaptive Reasoning Strategy.}
Unlike E5, H-STAR uses SQL to support semantic reasoning for quantitative tasks while relying solely on semantic reasoning for lexical queries. In contrast, ReAct interleaves reasoning and actions without adapting to task type. H-STAR’s hybrid approach, combining SQL logic and text understanding, excels at handling diverse tasks, outperforming ReAct on table reasoning tasks.

\section{Input Table Format}
\label{sec:input_format}
The input table format changes depending on the type of reasoning used. Since \AlgName uses a hybrid approach, the input table format also varies depending on the type of reasoning used. Besides the question and the in-context examples, the table is accompanied by a list of columns and a table caption if available. Providing more context aids in better semantic understanding \cite{singha2023tabular, sui2024table}.

\subsection{Textual Reasoning}
We convert the tables into linear, sequential text. Continuing with the format of previous textual reasoning methods \cite{ye2023large, chen2023large,wang2023chain}, we adopt the PIPE encoding i.e. plain text separated by `|'. Furthermore, we append the table with the caption and the list of columns similar to the format in Section \ref{sec:sr}. Example of the input table:\\ \\
{\small
{\fontfamily{qcr}\selectfont 
table caption: 2012–13 Exeter City F.C. season \\
/ \\
col : name | league | total \\
row 1: danny coles | 3 | 3 \\
row 4: john o'flynn | 11 | 12 \\
row 8: jamie cureton | 20 | 20 \\
*/ \\
columns: ['name', 'league', 'total']
}}

\subsection{Symbolic Reasoning} \label{sec:sr}We adopt the table prompt format from previous SQL-based methods Text-to-SQL \cite{rajkumar2022evaluating, cheng2022binding, nahid2024tabsqlify}. We include (1) the table schema containing {\fontfamily{qcr}\selectfont CREATE TABLE} followed by the schema, (2) the table header and all the rows separated by tabs, and (3) the list of columns along with the corresponding query. If the prompt exceeds the context limit, we truncate the table rows to fit within the limit. Example of the input table:\\ \\
{\small
{\fontfamily{qcr}\selectfont
CREATE TABLE 2012–13 Exeter City F.C. season( \\ 
    \indent row\_id int,\\
    \indent name text, \\
    \indent    league int, \\
    \indent    total int) \\
/ \\
All rows of the table: \\
SELECT * FROM w; \\
row\_id \indent name \indent league \indent total \\
1  \indent danny coles \indent 3 \indent 3 \\
4  \indent john o'flynn \indent 11 \indent 12 \\
8  \indent jamie cureton \indent 20 \indent 20 \\
/ \\
columns: ['name', 'league', 'total']
}}

\section{Implementation Details}
\label{sec: implm_details}
We provide the implementation details including the hyperparameters and prompt details for each of the individual steps. \AlgName operates in two key stages: (1) Table Extraction and (2) Adaptive Reasoning. The Table Extraction phase includes column extraction (SQL-based: {\fontfamily{qcr}\selectfont col\textsubscript{sql}} and text-based: {\fontfamily{qcr}\selectfont col\textsubscript{text}}) and row extraction (SQL-based: {\fontfamily{qcr}\selectfont row\textsubscript{sql}} and text-based: {\fontfamily{qcr}\selectfont row\textsubscript{text}}). The Adaptive Reasoning phase consists of SQL-based ({\fontfamily{qcr}\selectfont f\textsubscript{sql}}) and text-based ({\fontfamily{qcr}\selectfont f\textsubscript{text}}) reasoning components.

\subsection{Input Prompts}
\begin{figure*}[!htbp]
\centering
  \includegraphics[width=\linewidth]{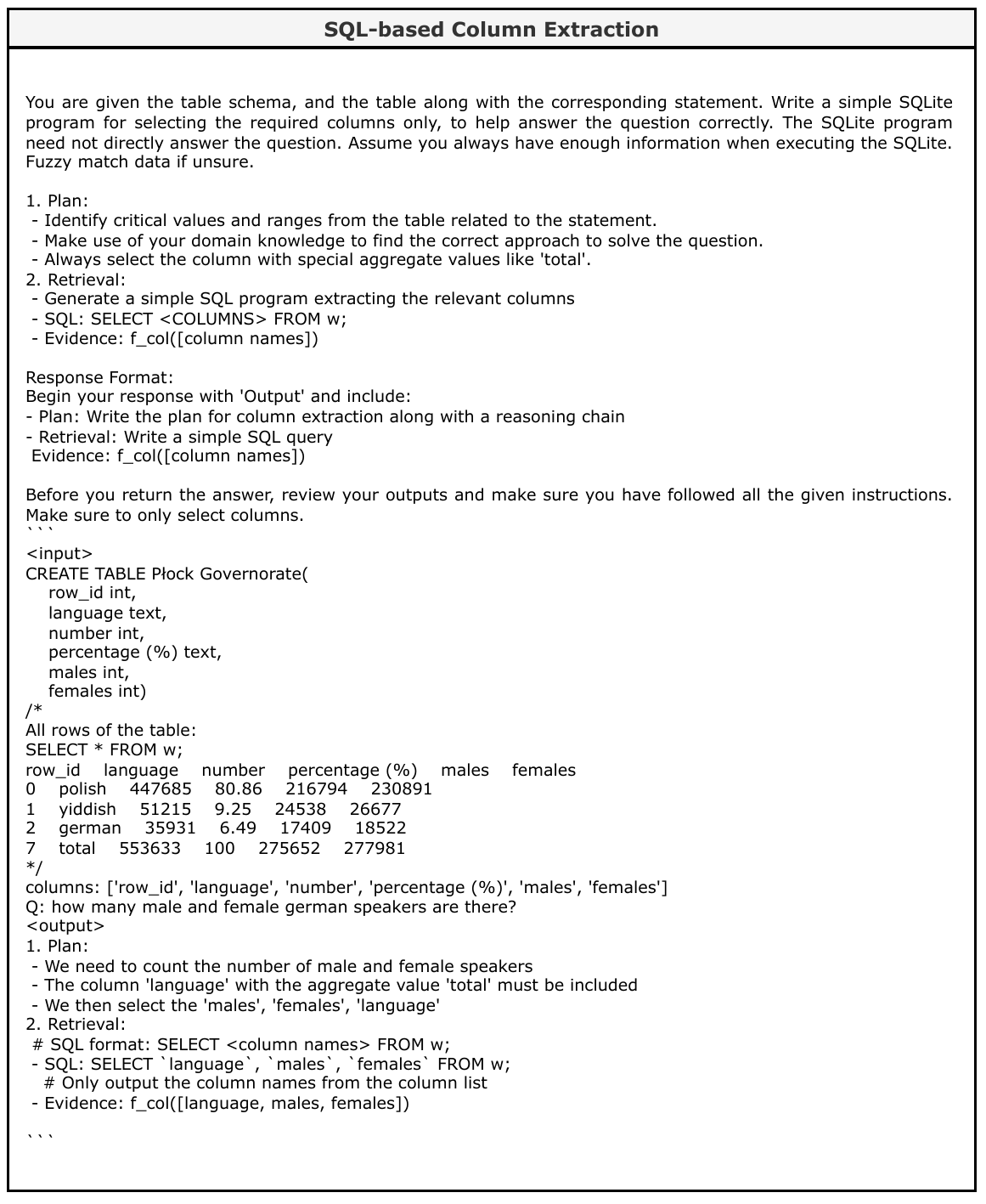}
  \caption{\small Prompt for SQL-based column extraction ({\fontfamily{qcr}\selectfont col\textsubscript{sql}}).}
  \label{fig:col_sql}
\end{figure*}

\begin{figure*}[!htbp]
\centering
  \includegraphics[width=\linewidth]{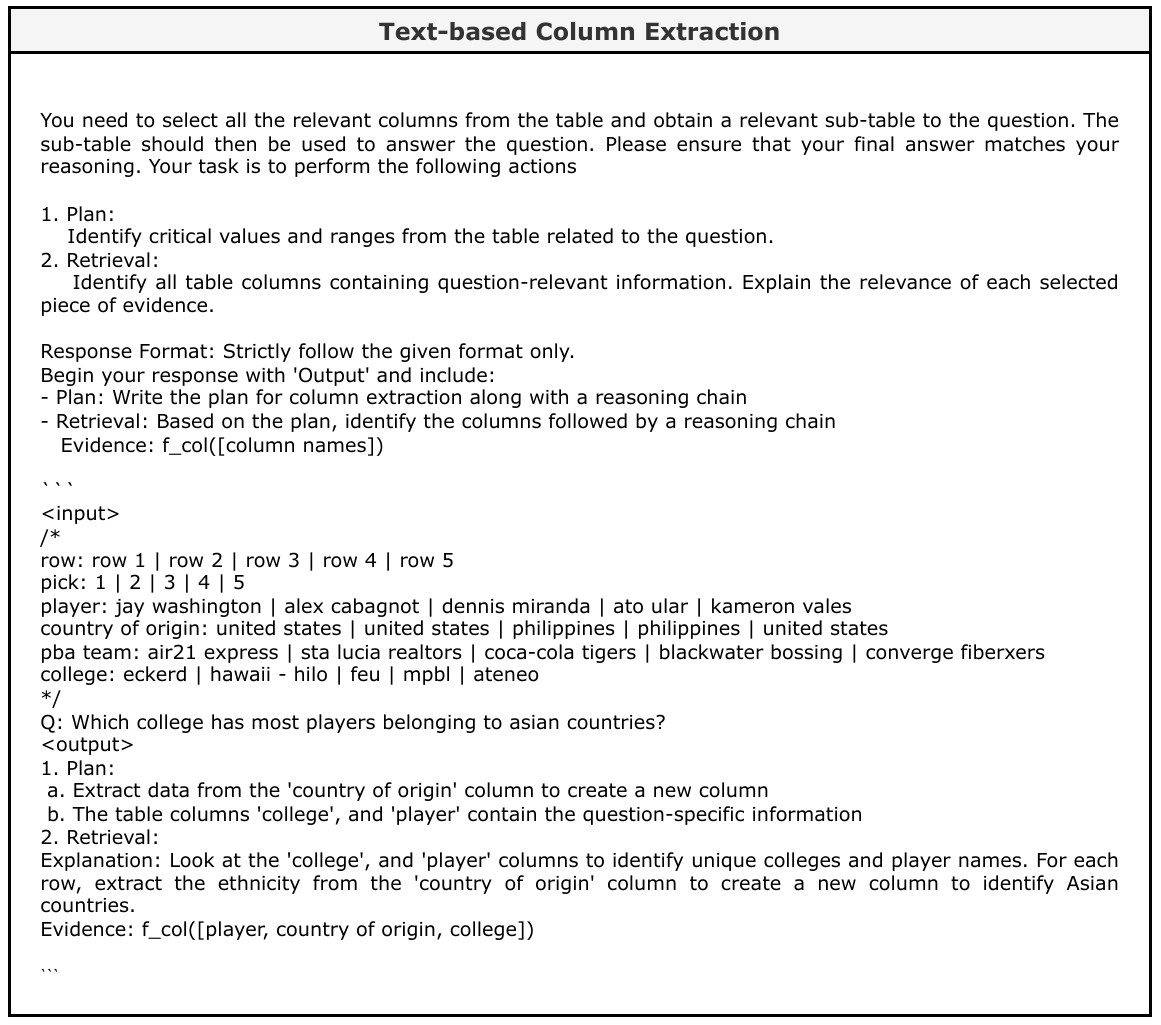}
  \caption{\small Prompt for Column-based column verification ({\fontfamily{qcr}\selectfont col\textsubscript{text}}).}
  \label{fig:col_text}
\end{figure*}

\begin{figure*}[!htbp]
\centering
  \includegraphics[width=\linewidth]{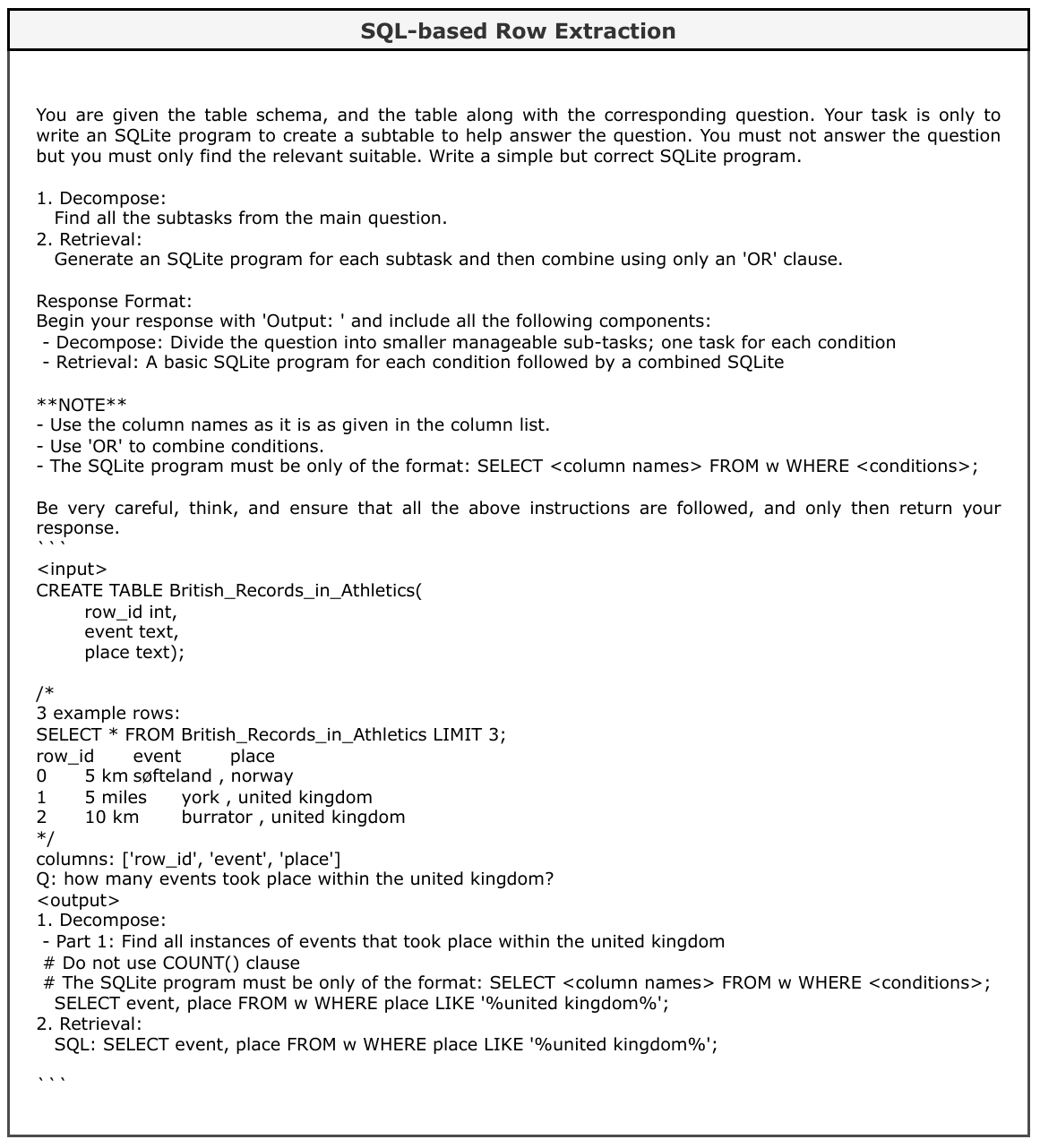}
  \caption{\small Prompt for SQL-based row selection ({\fontfamily{qcr}\selectfont row\textsubscript{sql}}).}
  \label{fig:row_sql}
\end{figure*}

\begin{figure*}[!htbp]
\centering
  \includegraphics[width=\linewidth]{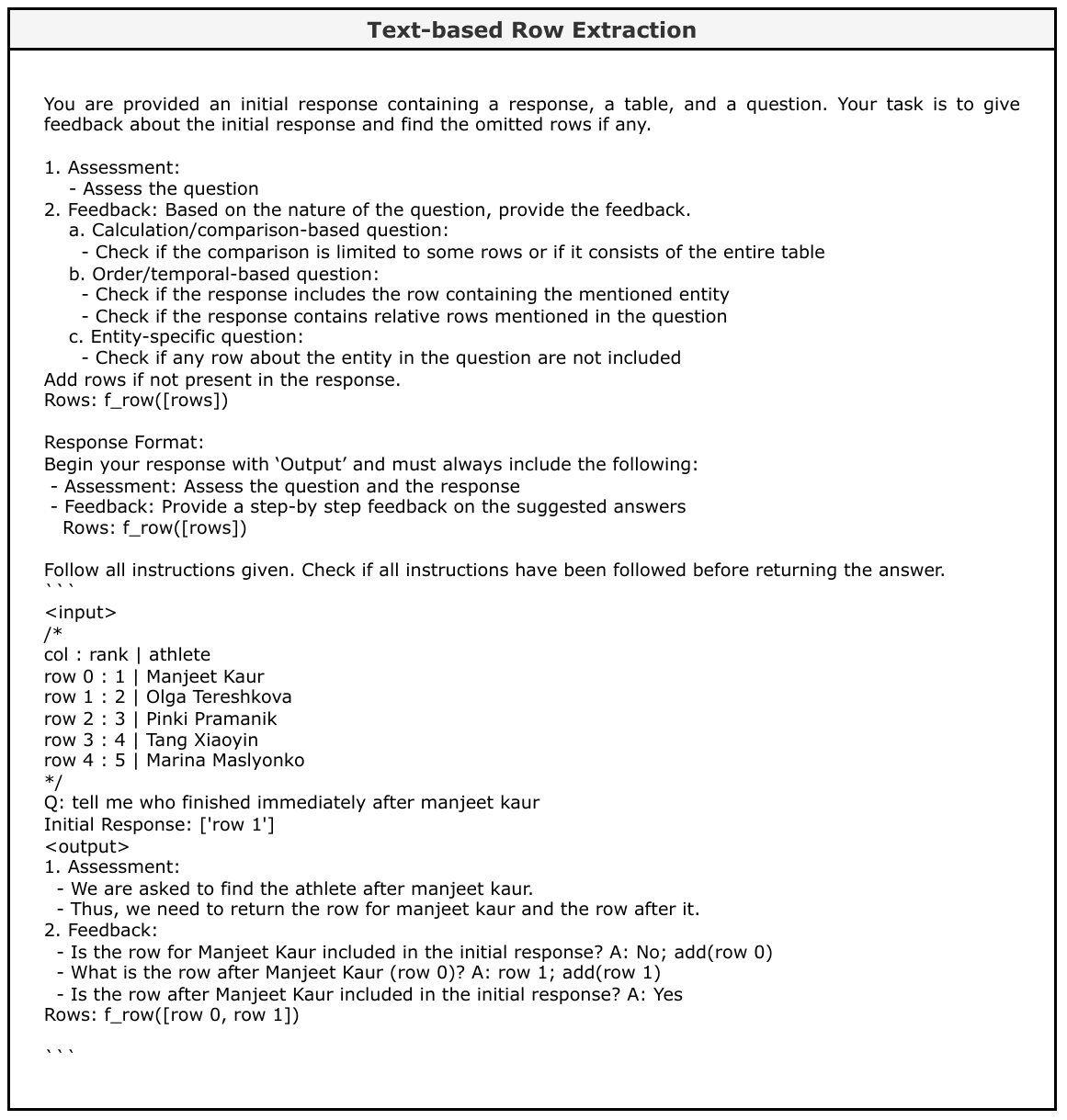}
  \caption{\small Prompt for text-based row verification ({\fontfamily{qcr}\selectfont row\textsubscript{text}}).}
  \label{fig:row_text}
\end{figure*}

\begin{figure*}[!htbp]
\centering
  \includegraphics[width=\linewidth]{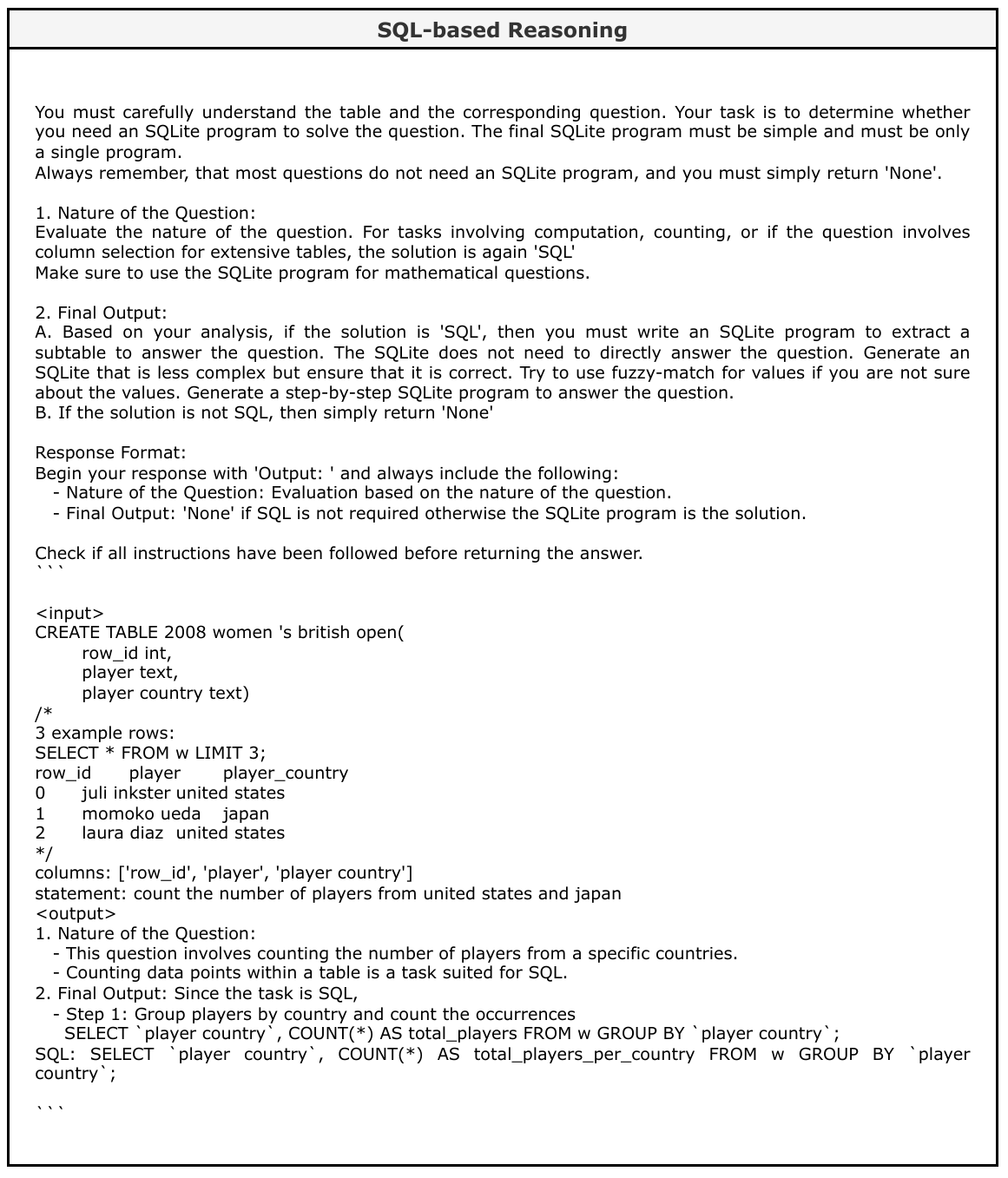}
  \caption{\small Prompt for SQL-based reasoning ({\fontfamily{qcr}\selectfont f\textsubscript{sql}}).}
  \label{fig:f_sql}
\end{figure*}

\begin{figure*}[!htbp]
\centering
  \includegraphics[width=\linewidth]{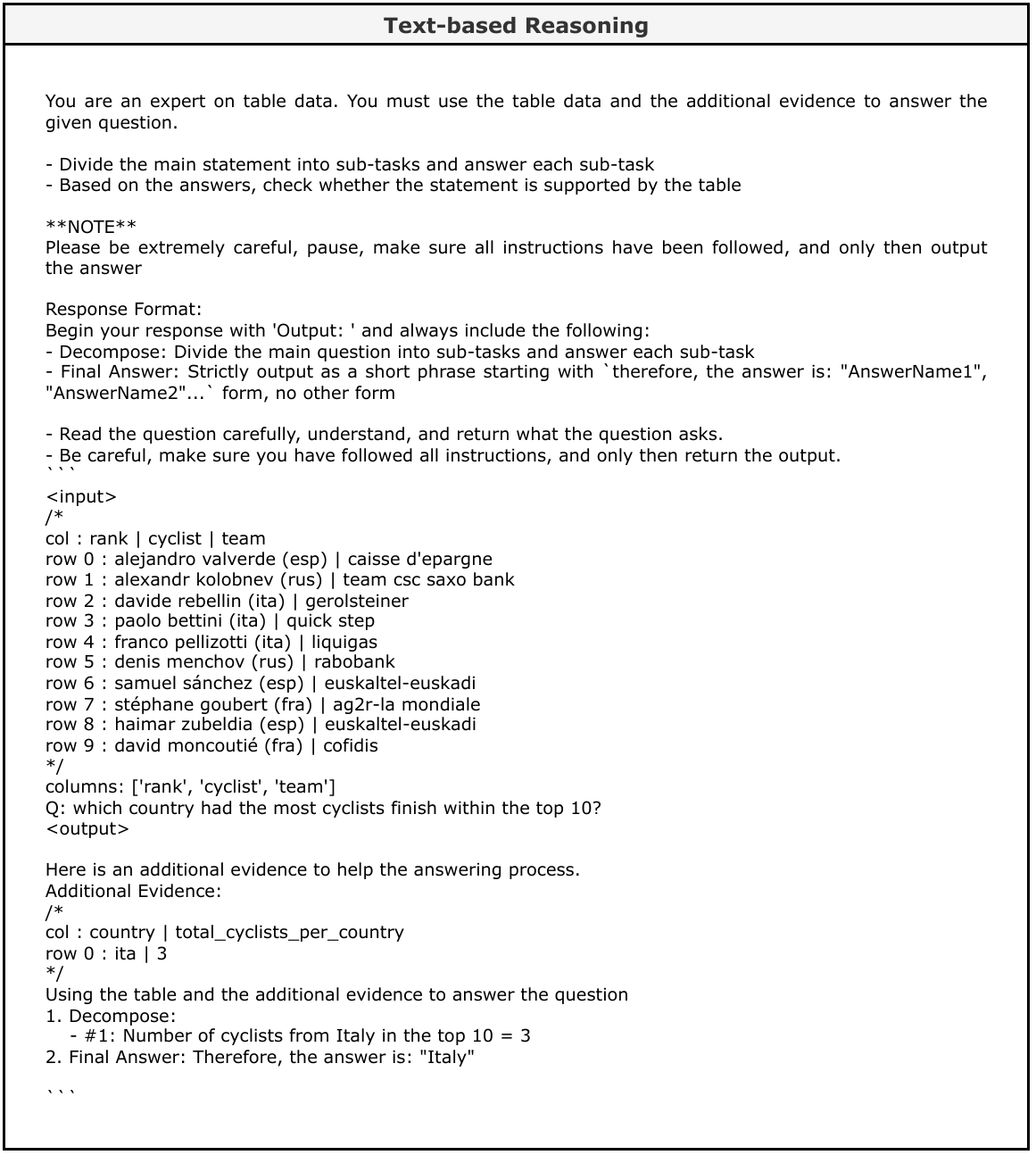}
  \caption{\small Prompt for text-based reasoning ({\fontfamily{qcr}\selectfont f\textsubscript{text}}).}
  \label{fig:f_text}
\end{figure*}

Figures \ref{fig:col_sql}, \ref{fig:col_text}, \ref{fig:row_sql}, \ref{fig:row_text}, \ref{fig:f_sql} and \ref{fig:f_text} illustrate the input prompts containing sample in-context learning examples for the steps {\fontfamily{qcr}\selectfont col\textsubscript{sql}}, {\fontfamily{qcr}\selectfont col\textsubscript{text}}, {\fontfamily{qcr}\selectfont row\textsubscript{sql}}, {\fontfamily{qcr}\selectfont row\textsubscript{text}}, {\fontfamily{qcr}\selectfont f\textsubscript{sql}}, and {\fontfamily{qcr}\selectfont f\textsubscript{text}} respectively.

\subsection{Hyperparameters}
Table \ref{tab:palm_parameter_settings} provides the details of PaLM-2 hyper-parameters used for the WikiTQ and TabFact datasets. Table \ref{tab:gpt_parameter_settings} showcases the hyperparameter setting for GPT-3.5-Turbo. The hyperparameters `samples' indicate the number of outputs taken for each step of the pipeline whereas the `examples' indicate the number of few-shot demonstrations for each step.

\begin{table}[!htbp]
\small
    \centering
    \setlength{\tabcolsep}{2.5pt}
    \scalebox{0.8}{
    \begin{tabular}{lcccccc}
        \toprule
        \multirow{1}{*}{\textbf{Function}} & \multicolumn{1}{c}{\textbf{temperature}} & \multicolumn{1}{c}{\textbf{top\_p}} & \multicolumn{1}{c}{\textbf{output\_tokens}} & \multicolumn{1}{c}{\textbf{samples}} & \multicolumn{1}{c}{\textbf{examples}} \\ \midrule
        \multicolumn{6}{c}{\textbf{WikiTQ}}\\
        \midrule
        {\fontfamily{qcr}\selectfont col\textsubscript{sql}} & 0.4 & 1.0 & 512 & 2 & 3 \\ \midrule
        {\fontfamily{qcr}\selectfont col\textsubscript{text}} & 0.7 & 1.0 & 512 & 2 & 3 \\ \midrule
        {\fontfamily{qcr}\selectfont row\textsubscript{sql}} & 0.4 & 1.0 & 512 & 2 & 3 \\ \midrule
        {\fontfamily{qcr}\selectfont row\textsubscript{text}} & 0.7 & 1.0 & 512 & 2 & 2 \\ \midrule
        {\fontfamily{qcr}\selectfont f\textsubscript{sql}} & 0.1 & 1.0 & 512 & 1 & 3 \\ \midrule
        {\fontfamily{qcr}\selectfont f\textsubscript{text}} & 0.0 & 1.0 & 512 & 1 & 4 \\ 
        \midrule
        \multicolumn{6}{c}{\textbf{TabFact}}\\
        \midrule
        {\fontfamily{qcr}\selectfont col\textsubscript{sql}} & 0.4 & 1.0 & 512 & 2 & 4 \\ \midrule
        {\fontfamily{qcr}\selectfont col\textsubscript{text}} & 0.7 & 1.0 & 512 & 2 & 3 \\ \midrule
        {\fontfamily{qcr}\selectfont row\textsubscript{sql}} & 0.4 & 1.0 & 512 & 2 & 4 \\ \midrule
        {\fontfamily{qcr}\selectfont row\textsubscript{text}} & 0.7 & 1.0 & 512 & 2 & 3 \\ \midrule
        {\fontfamily{qcr}\selectfont f\textsubscript{sql}} & 0.1 & 1.0 & 512 & 1 & 3 \\ \midrule
        {\fontfamily{qcr}\selectfont f\textsubscript{text}} & 0.0 & 1.0 & 512 & 1 & 5 \\ \bottomrule
    \end{tabular}}
    \caption{\small Hyperparameter settings for PaLM-2.}
    \label{tab:palm_parameter_settings}
\end{table}

\begin{table}[!htbp]
\small
    \centering
    \setlength{\tabcolsep}{2.5pt}
    \scalebox{0.8}{
    \begin{tabular}{lcccccc}
        \toprule
        \multirow{1}{*}{\textbf{Function}} & \multicolumn{1}{c}{\textbf{temperature}} & \multicolumn{1}{c}{\textbf{top\_p}} & \multicolumn{1}{c}{\textbf{output\_tokens}} & \multicolumn{1}{c}{\textbf{samples}} & \multicolumn{1}{c}{\textbf{examples}} \\  \midrule
        \multicolumn{6}{c}{\textbf{WikiTQ}}\\
        \midrule
        {\fontfamily{qcr}\selectfont col\textsubscript{sql}} & 0.3 & 1.0 & 512 & 2 & 3 \\ \midrule
        {\fontfamily{qcr}\selectfont col\textsubscript{text}} & 0.4 & 1.0 & 512 & 2 & 2 \\ \midrule
        {\fontfamily{qcr}\selectfont row\textsubscript{sql}} & 0.3 & 1.0 & 512 & 2 & 3 \\ \midrule
        {\fontfamily{qcr}\selectfont row\textsubscript{text}} & 0.4 & 1.0 & 512 & 2 & 3 \\ \midrule
        {\fontfamily{qcr}\selectfont f\textsubscript{sql}} & 0.1 & 1.0 & 512 & 1 & 4 \\ \midrule
        {\fontfamily{qcr}\selectfont f\textsubscript{text}} & 0.0 & 1.0 & 512 & 1 & 4 \\ 
        \midrule
        \multicolumn{6}{c}{\textbf{TabFact}}\\
        \midrule
        {\fontfamily{qcr}\selectfont col\textsubscript{sql}} & 0.2 & 1.0 & 512 & 2 & 4 \\ \midrule
        {\fontfamily{qcr}\selectfont col\textsubscript{text}} & 0.4 & 1.0 & 512 & 2 & 3 \\ \midrule
        {\fontfamily{qcr}\selectfont row\textsubscript{sql}} & 0.4 & 1.0 & 512 & 2 & 4 \\ \midrule
        {\fontfamily{qcr}\selectfont row\textsubscript{text}} & 0.5 & 1.0 & 512 & 2 & 3 \\ \midrule
        {\fontfamily{qcr}\selectfont f\textsubscript{sql}} & 0.1 & 1.0 & 512 & 1 & 4 \\ \midrule
        {\fontfamily{qcr}\selectfont f\textsubscript{text}} & 0.0 & 1.0 & 512 & 1 & 5 \\ \bottomrule
    \end{tabular}}
    \caption{\small Hyperparameter settings for GPT-3.5-Turbo.}
    \label{tab:gpt_parameter_settings}
\end{table}

\section{Error Analysis: Case Study}
\label{sec:row_col}
\begin{figure*}[!htbp]
  \centering
    \includegraphics[width=\linewidth]{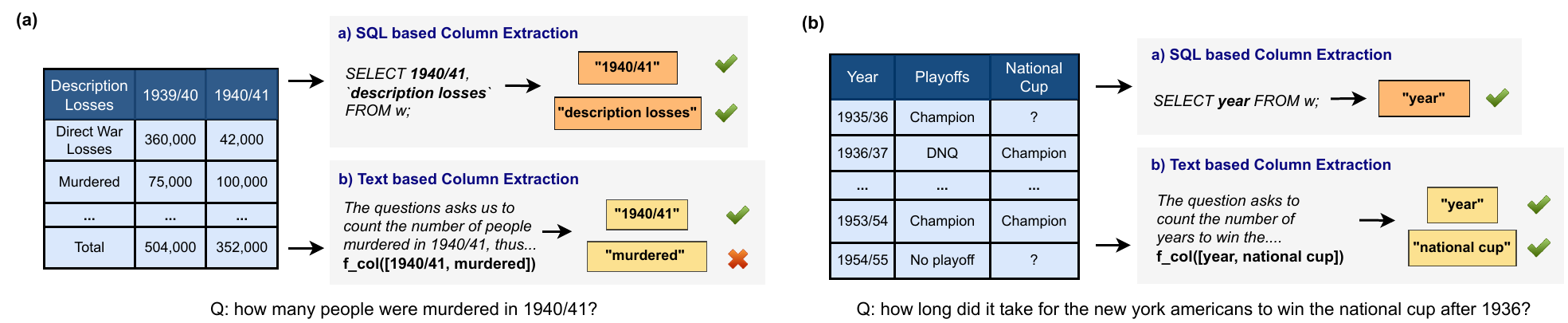}
    \caption{\small Comparison of SQL-based and text-based column extraction methods for answering table queries: (a) SQL-based selection accurately identifies the columns ("1940/41", "description losses"), while text-based selection mistakenly includes "murdered" as a table column. (b) The text-based method selects both required columns ("year" and "national cup"), while the SQL-based approach overlooks "national cup" as a requirement.}
  \label{fig:column}
\end{figure*}

\begin{figure*}[!htbp]
\centering
  \includegraphics[width=\linewidth]{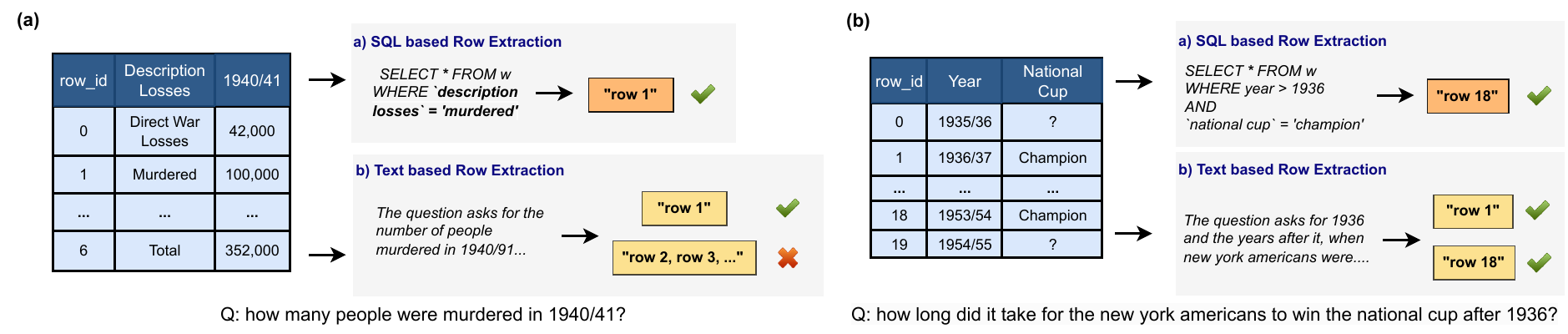} 
  \caption{\small Comparison of SQL-based and text-based row extraction methods for answering table queries. (a) SQL-based selection accurately selects the necessary rows ("Murdered"), while text-based selection incorrectly includes additional rows; (b) Text-based method correctly selects relevant rows ("1953/54", "1936/37"), while SQL-based selection misinterprets the query, including only the row for the year "1953/54".} 
  \label{fig:row}
  \end{figure*}

\begin{figure*}[!htb]
\centering
  \includegraphics[width=\linewidth]{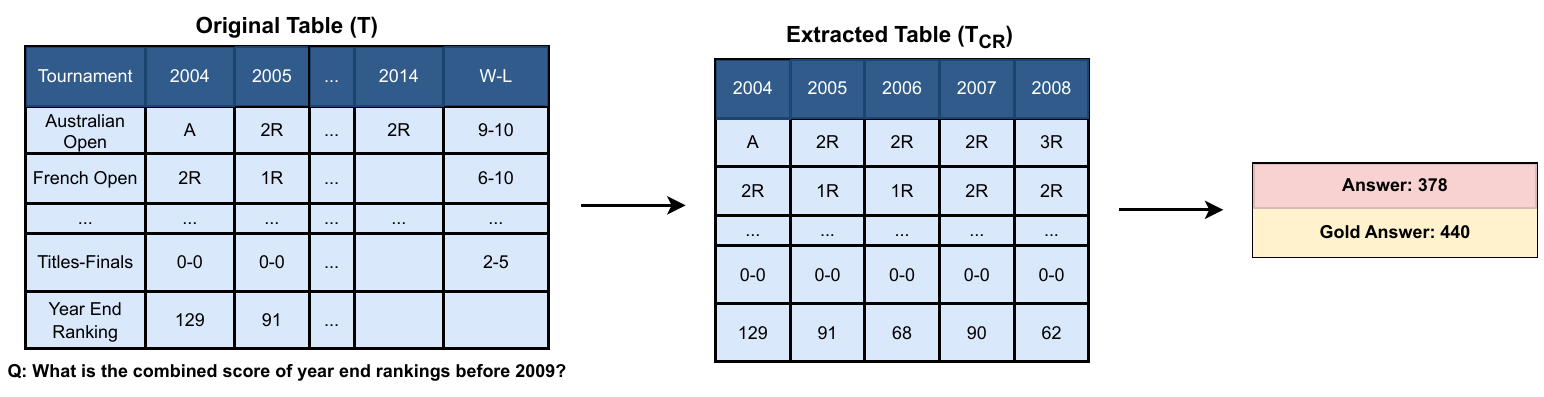}
  \caption{\small \textbf{Case Study: `Missing Columns'.} An illustration of the error `Missing Columns'. In the above example, the extracted table does not contain the column `Tournament' that contains the `Year End Ranking'. The omission of the `Tournament' column affects the downstream steps in the pipeline leading to an insufficient row extraction and reasoning.}
  \label{fig:column_error}
\end{figure*}

\begin{figure*}[!htbp]
\centering
  \includegraphics[width=\linewidth]{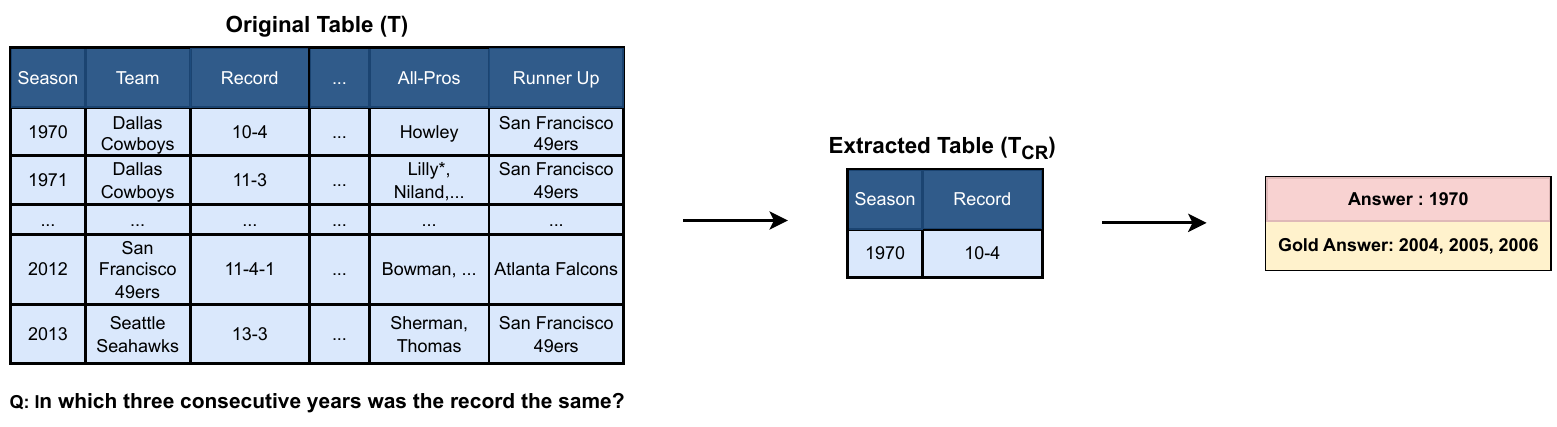}
  \caption{\small \textbf{Case Study: `Missing Rows'.} An illustration of the error `Missing Rows'. From the above example, it can be seen that while the extracted columns are correct, the row extraction is incorrect. Thus, as the extracted table lacks the necessary information, the final answer is incorrect.}
  \label{fig:row_error}
\end{figure*}

\begin{figure*}[!htbp]
\centering
  \includegraphics[width=\linewidth]{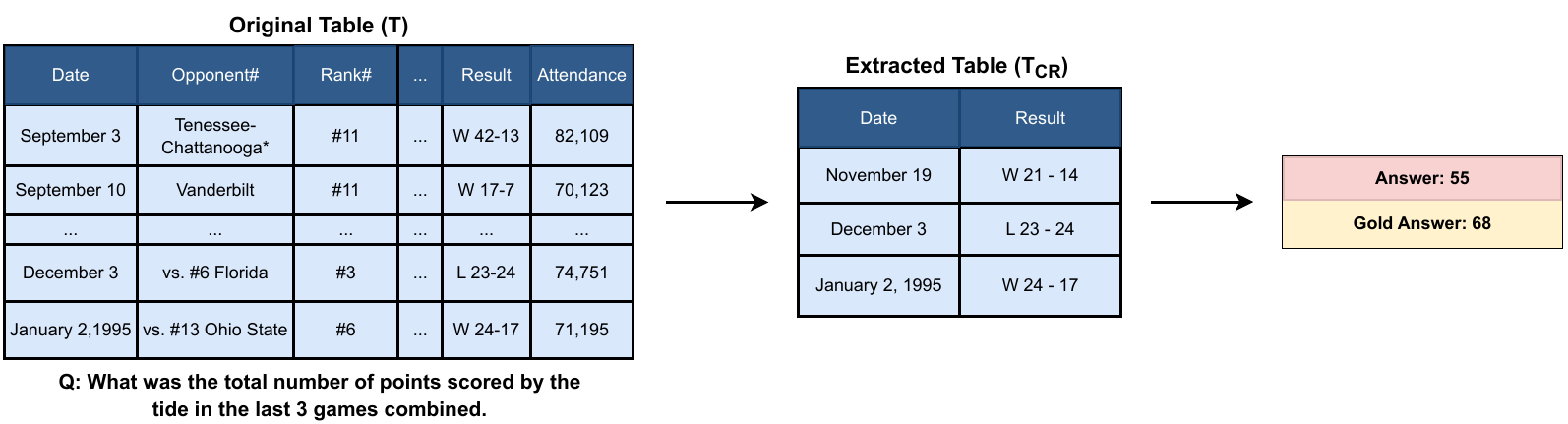}
  \caption{\small \textbf{Case Study: `Incorrect Reasoning'.} An illustration of the error `Incorrect Reasoning'. Despite the correct question-relevant table being extracted, the model misinterprets the question. It adds up the scores by the opposing team instead thus returning an incorrect answer.}
  \label{fig:reason_error}
\end{figure*}

\begin{figure*}[!htbp]
\centering
  \includegraphics[width=\linewidth]{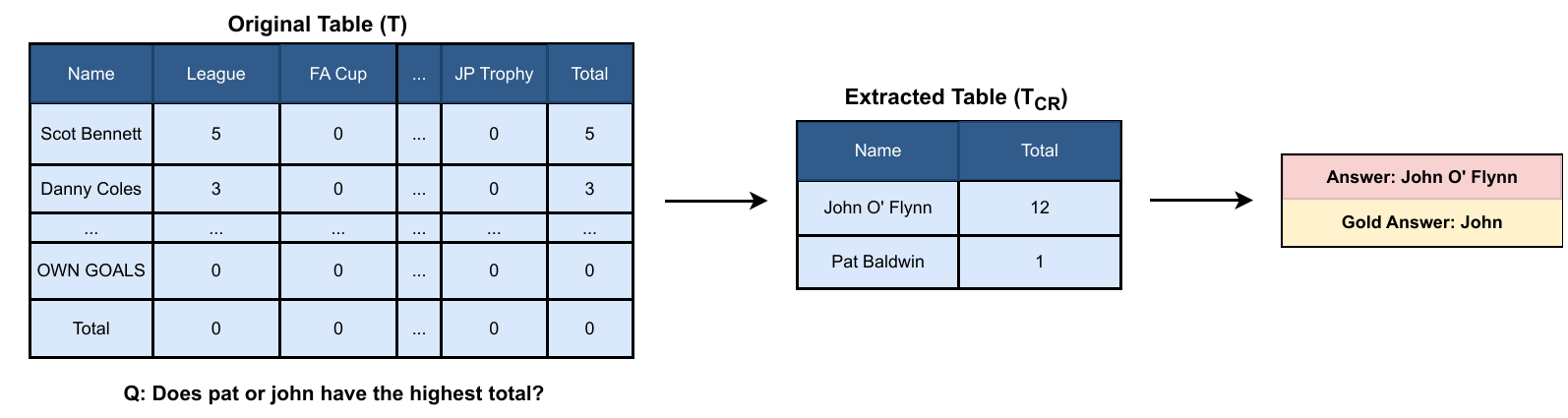}
  \caption{\small \textbf{Case Study: `Incorrect Annotation'.} An illustration of the error `Incorrect Annotation'. In the example, the query-specific table is correctly extracted and the final reasoning leads to the correct answer. However, the prediction is penalized for not being an `exact match'.}
  \label{fig:annotation_error}
\end{figure*}

Figures \ref{fig:column} and \ref{fig:row} showcase how semantic and symbolic approaches can individually fail for the column and row extraction steps. Figures \ref{fig:column_error}, \ref{fig:row_error}, \ref{fig:reason_error} and \ref{fig:annotation_error} illustrate error cases for each individual step within \AlgName.

\newpage

\clearpage

\end{document}